# Unravelling magnetic vortex-like excitations through rapid thermal quenching in low - carbon steel

P. C. Mahato[1], Suprotim Saha[1], Ritesh Kumar[1], D. Banik[2], K. Mondal[2], Shyam Kumar Choudhary[3], Ashish Garg[4], S. S. Banerjee[1,+]

[1]*Department of Physics, Indian Institute of Technology Kanpur, Kanpur, Uttar Pradesh 208016, India*

[2]*Department of Material Science and Engineering, Indian Institute of Technology Kanpur, Kanpur, Uttar Pradesh 208016, India*

[3]*Graphene Center, Tata Steel Limited, Jamshedpur, Jharkhand 831007, India.*

[4] *Department of Sustainable Energy Engineering, Kotak School of Sustainability, Indian Institute of Technology Kanpur, Kanpur, Uttar Pradesh 208016, India*

Corresponding author's email ID:

+: satyajit@iitk.ac.in

**Abstract**

Steel, traditionally valued for its structural strength, emerges in this study as a remarkable material for exploring novel magnetic phenomena. We investigate how common processing techniques—thermal treatments and mechanical strain—significantly affect the magnetic properties of low-carbon steels (0.05% by weight). Our findings show that slow annealing enlarges the grain size, enhancing magnetic susceptibility, while rapid quenching reduces grain size, resulting in a decreased magnetic response. Quenching low-carbon steel produces significant increase in the fraction of high-angle grain boundaries and a rapid spatial variation of local magnetic anisotropy between grains, a feature which is unachievable with mechanical straining even up to the material's ultimate tensile strength. Tensile-straining of low-carbon steel enhances magnetic susceptibility through altered magnetic anisotropy, contrary to the observed decrease of susceptibility in quenched low-carbon steel. Magnetic force microscopy and micromagnetic modelling of our data reveal that, the reduced magnetic susceptibility in quenched steel is a result of the presence of intriguing magnetic excitations akin to magnetic vortices. These localized structures act as strong magnetic domain wall pinning centres, causing the observed decrease in magnetic susceptibility in these quenched low-carbon steels. Beyond its established structural utility, low-carbon steel combines mechanical stability with favourable magnetic properties, positioning it as a strong platform for magnetic device applications.

## 1. Introduction

Magnetization dynamics in magnetic materials are widely studied. Weiss's molecular field theory introduced magnetic domains, determined by the interplay of magneto-crystalline anisotropy and exchange energy [1-3]. Such idealized models are complicated by the presence of defects, like dislocations and voids, in real magnetic systems [4-6]. Studies show how domain walls (DW) interact with defect sites which act as pinning centres [7-10], hindering DW motion under an external magnetic field. Such irregular DW pinning and de-pinning are manifested as discontinuous steps in magnetization in Barkhausen noise studies [11-13]. Magnetic properties are also sensitive to thermal and stain-induced magneto-mechanical effects [14-18]. It is well known that strain-induced changes impact dislocation density, DW pinning, and magnetization hysteresis [19]. Therefore, a material's magnetic properties often reflect its imperfections—whether chemical, structural, or strain-induced [20]. The spatial variation in microscopic crystallographic description of the material, in particular, misorientations, defects, and extended defects such as grain boundaries (GB), influences its magnetic anisotropy and in turn, the local magnetization landscape. Condensed matter physics offers numerous examples of how grain-boundaries and misorientations influence material properties, for instance, enhanced local magnetic moments in the GB region for ferrite and other Fe-based magnets [21], the dependence of critical current on local misorientation in superconductors [22], Josephson vortices and spontaneous half-integer vortices at the GBs and tricrystal junctions [23, 24], to name a few. Recent research themes in magnetism attempt at discovering new quantum materials exhibiting a correlation between microstructure, strong electronic correlations and strain effects, driving unique magnetic properties [5, 25] and creating exotic magnetic excitations like magnetic vortices [26], skyrmions [27] and others [28-32]. In particular, magnetic vortices, which are local swirls of magnetic moments stabilized by a complex interplay of exchange and dipolar interactions, hold significant potential for device applications [33]. Although diverse range of materials is available, the potential of steel to host exotic and tunable magnetic ground states [34] remains under-explored. This is noteworthy because steel's mechanical and metallurgical properties are well understood, and its processing routes are finely tuned and widely established [35].

Steel, an alloy of iron (Fe) and carbon (C), along with elements like Mn, P, Cr and Si, is vital to modern engineering due to its strength, durability, and adaptability. Its microstructure, governed by carbon content and processing temperatures, comprises phases such as Ferrite (soft, ductile BCC phase with low carbon), Austenite (tough FCC phase with higher carbon, usually stable at high temperatures), and Cementite, $Fe_3C$ (hard, brittle phase with 6.67 wt% carbon), see Fig. S1(a) in SI-1 (supplementary information) for a detailed Fe-C phase-diagram [36]. This is an active area of research even recently [37] – indicating the richness of steel's phase diagram in terms of crystallography, microstructures and magnetic properties. Tailoring these phases via thermal treatment and alloying optimizes steel for varied applications.

While mainstream metallurgical studies often focus on steel's magnetic properties—like coercivity, saturation magnetization, and Barkhausen noise—for non-destructive evaluation of steel-quality [14, 38-47]; our experiments on low-carbon steels (0.05% by weight) reveal that thermal treatment and mechanical strain significantly influence their magnetic behaviour. Such thermal and mechanical treatment have a major impact on the grain size and grain distributions of steel. Dynamic observations by Kerr and Lorentz microscopy reveal that high-angle grain boundaries (HAGB) have a more significant effect on magnetic domain structures than low-angle grain boundaries (LAGB)[48, 49]. We found that rapid thermal quenching in low-carbon (0.05%) steel significantly reduces the grain-size and produces a substantial HAGB fraction, which is otherwise unachievable even with the highest level of tensile-straining, up to the material's ultimate tensile strength (UTS). Our structural characterization and simulations indicate that rapid thermal quenching generates a mosaic of grains with sharply varying local magnetic anisotropy, an unusual and difficult-to-engineer combination of microstructure and magnetic features, that strongly shapes the material's magnetic response. We find that the magnetic susceptibility ($\chi$) of the thermally quenched low-carbon steel drops, as compared to the as-received steel. Magnetic force microscopy (MFM)-based imaging reveals presence of localized magnetic structures akin to magnetic vortices in quenched low-carbon steel. Our micromagnetic modelling considers that quenched low-carbon steel possesses a mosaic of grains with rapid changes in local magnetic anisotropy across grains. With this micromagnetic model, we match our observed behaviour in χ, and we see the presence of swirls in local magnetization near the DW edges which act as pinning-centres for the DW expansion with magnetizing field. These swirls are very similar to the structures observed in our MFM studies.

## 2. Experimental Results and Discussions

### 2.1. Effects of thermal treatment on low-carbon (0.05%) steel

Our study primarily focuses on the effects of different thermal treatments on the magnetic behaviour of a low-carbon (0.05% by weight, see the complete elemental composition in table S1, SI-1) cold-rolled steel sheets (labelled $S^{0.05\%}$). Samples with dimensions 5 mm × 5 mm × 0.8 mm, taken from the as-received (AR) cold-rolled $S^{0.05\%}$ steel sheets, are designated as $S^{0.05\%}_{AR}$. The $S^{0.05\%}_{AR}$ is subjected to: (a) furnace-annealing (FA) after Austenitization at 1000°C, holding for 2 minutes followed by slow furnace-cooling to room temperature over ~ 20 hours ($S^{0.05\%}_{FA}$); and (b) rapid water-quenching (WQ) after complete Austenitization for the same temperature and time ($S^{0.05\%}_{WQ}$), see Fig. S1(b) in SI-1 for the Time-temperature-transformation (TTT) and continuous cooling transformation (CCT) diagrams for $S^{0.05\%}$ steel [50] along with our heat treatment regimes. The crystallographic and morphological properties of $S^{0.05\%}_{AR}$, $S^{0.05\%}_{FA}$ and $S^{0.05\%}_{WQ}$ samples were studied using electron backscattered diffraction (EBSD) and X-ray diffraction (XRD), see SI-2 and SI-3 for details. XRD spectra is similar for $S^{0.05\%}_{AR}$, $S^{0.05\%}_{FA}$ and $S^{0.05\%}_{WQ}$ samples (Fig. 1(a)), implying dominant ferrite content in all 3 samples. Since the carbon content is very low (0.05% by weight), there is hardly any shift or splitting in the [1 1 0] ferrite-plane ($2\theta \sim 52.3°$ for Co-$K_\alpha$, see Fig. 1(a)) [51] even after quenching. The Inverse pole figure (IPF) maps obtained from EBSD in Figs. 1(b)-(d) show the polycrystalline grain orientations in the $S^{0.05\%}_{AR}$, $S^{0.05\%}_{FA}$ and $S^{0.05\%}_{WQ}$ samples, respectively.

In $S^{0.05\%}_{FA}$, we see annealing increases grain size w.r.to $S^{0.05\%}_{AR}$steel (compare Figs. 1(b) and (c)). In $S^{0.05\%}_{WQ}$, quenching reduces grain size significantly (compare Figs. 1(b) and 1(d)) and introduces a wider variation in grain orientation (note the change in colour across grains in Fig. 1(d)). From the IPF maps (Figs. 1(b)-(d)), the mean grain-size of $\mathrm{S}_{\mathrm{FA}}^{0.05\%}$ ~ (43.37 $\pm$ 0.92) μm, $\mathrm{S}_{\mathrm{AR}}^{0.05\%}$ ~ (23.33 $\pm$ 0.77) μm, and $\mathrm{S}_{\mathrm{WQ}}^{0.05\%}$ ~ (14.38 $\pm$ 0.34) μm were obtained. See the quantitative comparison of the average grain-sizes in Table S4, SI-3, determined from Gaussian fits to the statistical distribution of the grain-sizes (Fig. S2(g)-(i)) of $\mathrm{S}_{\mathrm{AR}}^{0.05\%}$, $\mathrm{S}_{\mathrm{FA}}^{0.05\%}$ and $\mathrm{S}_{\mathrm{WQ}}^{0.05\%}$ respectively (also see the optical (Fig. S2(a)-(c)) and SEM micrographs (Fig. S2(d)-(f) of $\mathrm{S}_{\mathrm{AR}}^{0.05\%}$, $\mathrm{S}_{\mathrm{FA}}^{0.05\%}$ and $\mathrm{S}_{\mathrm{WQ}}^{0.05\%}$ samples). Generally, finer grains enhance toughness, hardness, and shock resistance compared to coarser-grained steel [52, 53]. This agrees with our Vickers hardness test result which shows ~100% increase in hardness of $\mathrm{S}_{\mathrm{WQ}}^{0.05\%}$ compared to $\mathrm{S}_{\mathrm{AR}}^{0.05\%}$ steel and a drop in hardness for $\mathrm{S}_{\mathrm{FA}}^{0.05\%}$ steel (see Figs. S3(a)-(c), and table S5, SI-4).

Figs. 2(a)-(c) show the grain boundary maps of $\mathrm{S}_{\mathrm{AR}}^{0.05\%}$, $\mathrm{S}_{\mathrm{FA}}^{0.05\%}$ and $\mathrm{S}_{\mathrm{WQ}}^{0.05\%}$ samples respectively, with the high-angle grain boundaries (HAGB) coloured in blue (having grain-misalignment > 15°, see the colour code for grain boundaries in top-left corner of Fig. 2). Fig. 2(c) clearly shows that quenching results in a significant enhancement of HAGB fraction (the fraction of grain boundaries having misalignment >15° normalized by total GB length) as compared to $\mathrm{S}_{\mathrm{AR}}^{0.05\%}$ and $\mathrm{S}_{\mathrm{FA}}^{0.05\%}$. The HAGB fraction is 84.6% for $\mathrm{S}_{\mathrm{WQ}}^{0.05\%}$ compared to 14.3% and 23.8% for $\mathrm{S}_{\mathrm{AR}}^{0.05\%}$ , $\mathrm{S}_{\mathrm{FA}}^{0.05\%}$ respectively, see Fig. 2(d) and detailed quantification of HAGB and low-angle grain boundaries (LAGB) in Table S4, SI-3. Such high levels of HAGB fraction or local strain is unachievable by mechanical straining, even up to the material's UTS (discussed later), see Fig. 2(e). Fig. 2(c) and (d) collectively suggest that thermal quenching produces tiny and irregularly shaped crystallites (~ 14 μm, see table S4 in SI-3) with very high degree of crystallographic misalignment amongst them. Therefore, from the standpoint of magnetism, quenching produces a mosaic of rapidly changing local anisotropy direction from grain to grain. We discuss below the consequences of such thermal treatments on magnetic properties.

We investigate the high-frequency ($f \sim 72$ MHz) AC susceptibility ($\chi_{AC}$) response to focus on magnetization dynamics while avoiding the slow domain wall regime (kHz to sub-kHz) prevalent in steel due to grain boundary effects. Operating at high frequencies also minimizes $1/f$ noise and helps determine the sample's response to an oscillating magnetic field. We determine the $\chi_{AC}$ response of the $S^{0.05\%}_{AR}$, $S^{0.05\%}_{FA}$ and $S^{0.05\%}_{WQ}$ steel samples using a Tunnel Diode Resonator (TDR)-based susceptometer [54, 55]. We previously demonstrated [56] that a compact TDR-based susceptometer enables real-time monitoring of high-frequency $\chi_{AC}$ responses in steel, providing a practical tool for assessing structural integrity and predicting impending damage in steel-based infrastructures. Briefly, TDR is a self-resonant *LC* tank circuit which measures the $\chi_{AC}$ of a material from the shift in the circuit's resonant frequency $f_0(\sim\ 1/2\pi\sqrt{LC})$, $\chi_{AC}\ \propto \frac{\Delta f_m}{f_0}$ where $\Delta f_m$ is the shift in frequency because of the change in inductance *L* of the tank circuit by the material ($f_0 \sim 72$ MHz for our TDR circuit, for details see SI-5). Figures 2(f)-(g) show the TDR-based $\chi_{AC}$ response measured as a function of magnetic field (*H*) for the $S^{0.05\%}_{FA}$ and $S^{0.05\%}_{WQ}$ samples compared to the $S^{0.05\%}_{AR}$ sample. In the *H*-range studied (< 300 Oe) there is no significant variation in the $\chi_{AC}$ value. Therefore, any variations in $\chi_{AC}$ cannot be attributed to any differences in the thermomagnetic history of steel. Fig. 2(f) shows that the $\chi_{AC}$ of $S^{0.05\%}_{FA}$ is higher than $S^{0.05\%}_{AR}$ by ~ 11.7%. However, the $\chi_{AC}$ for $S^{0.05\%}_{WQ}$shows an opposite trend viz., $\chi_{AC}$ of $S^{0.05\%}_{WQ}$sample is lower than that of $S^{0.05\%}_{AR}$ by ~ 14.2 % (Fig. 2(g)). We also show that the fractional change from the DC magnetization response of these samples (Fig. S5(a)-(b) in SI-6) is in good agreement with the AC response (also see Table S4 in SI-3). In $S^{0.05\%}_{WQ}$samples, rapid cooling induces thermal contraction mismatches, dislocation generation, grain boundary stresses, and residual stresses, all of which may collectively influence the magnetic properties of these low-carbon steels. Below, we show that the high strain levels produced by quenching cannot typically be achieved through conventional tensile straining. Accordingly, we compare our magnetization results on thermally treated $S^{0.05\%}$ steel with those from the same batch subjected to controlled uniaxial tensile strains.

## 2.2. Studying pure strain effects on low-carbon (0.05%) steel

For pure strain effect study, the as-received $S^{0.05\%}$ steel sample isn't exposed to any thermal treatments. We apply a uniaxial tensile-load at a constant strain rate of $0.001 s^{-1}$ to the $S^{0.05\%}$ work-piece for stress (σ) - strain (ε) measurements (work-piece dimensions and relevant ASTM standards are detailed in SI-7, also see Fig. S6 in SI-7). The σ-ε curve for $S^{0.05\%}$ is shown in Fig. 3(a). The σ-ε curve (Fig. 3(a)) clearly shows the yield strength at ~ 340 MPa, expected for such low-carbon steels [57]. Based on the σ-ε plot, four different $S^{0.05\%}$ strained samples were prepared: sample $S_{TS_1}^{0.05\%}$ was strained within the elastic limit, up to $TS_1$ (see green circle marked $TS_1$ in Fig. 3(a)), sample $S_{TS_2}^{0.05\%}$ was strained to slightly above the elastic regime (exceeding yield strength) up to $TS_2$, the sample $S_{TS_3}^{0.05\%}$ was strained deep into the plastic regime near UTS up to $TS_3$, and sample $S_{TS_4}^{0.05\%}$ was strained beyond breaking stress till $TS_4$ (see the vertical drop beyond UTS in Fig. 3(a)).

After straining, sections were cut from each sample's central gauge length (for the fractured sample or $S_{TS_4}^{0.05\%}$, very close to the point of fracture), with dimensions 6 mm × 2 mm × 1.9 mm for EBSD to understand the effect of microscopic strain and eventually its imprint on magnetism. The highest tensile-strained sample $S_{TS_4}^{0.05\%}$ show a mosaic of grains with random orientations (Fig. 3(c)), quite similar in appearance to the quenched sample $S_{WQ}^{0.05\%}$ (Fig. 1(d)). Similar to Figs. 2(a)-(c), we generated grain boundary maps for the four strained samples (see Fig. S7(e)-(h) and Fig. 3(d)-(e)). The $S_{TS_1}^{0.05\%}$ sample (i.e., strained within elastic limit, see Fig. 3(d)) shows a HAGB fraction close to that of the $S_{AR}^{0.05\%}$ sample (~ 19 %, compare Figs. 2(d) and (e)). It is particularly noteworthy that even straining up to the material's breaking strength does not produce a HAGB fraction close to what is achieved by quenching ($S_{TS_4}^{0.05\%}$ has a HAGB fraction of ~ 31.9%, significantly lower than $S_{WQ}^{0.05\%}$ (~ 84.6%), see the comparison between Figs. 2(d)-(e)). Below, we further elaborate on the comparison of the microstructural details of the tensile-strained ($S_{TS_1}^{0.05\%} - S_{TS_4}^{0.05\%}$) steels from the view of microscopic analysis of low-angle misorientations.

Elastic strain usually results in changes in unit cell parameter(s). When the material is strained beyond the elastic limit, it undergoes permanent deformation and the distortions in the crystal lattice are relieved by the formation of dislocations[58, 59]. Plastic strain can be quantified by analyzing local crystallographic misorientation using a pixel-based analysis of IPF maps. We calculate the Kernel Average Misorientation (KAM) using KAM = $\frac{1}{4}\sum_{i=1}^{4} \theta\ (P, P_i)$, where $\theta(P, P_i)$ is the misorientation between point $P$ in a pixel and its four nearest neighbours on a square grid [59] in the IPF maps (see

Figs. 1(b)-(d) and 3(b)-(c)). For calculating KAM, a tolerance angle of 5° was set to exclude larger misorientations between neighbouring grains. In other words, only the grids having a misorientation less than this tolerance limit (5°) with their neighboring 4 grids are considered. KAM, therefore, gives a measure of the local lattice rotations. Inset of Fig. 3(f) shows log-normal probability distribution of KAM for two maximally strained samples, $S_{TS_3}^{0.05\%}$ and $S_{TS_4}^{0.05\%}$. Fig. 3(f) plots the mean KAM (<KAM>), derived from the log-normal fits for misorientation distribution of four strained sample, versus true macroscopic strain ($e = ln(1 + \varepsilon)$), revealing a clear increase in <KAM> with strain and supporting the effectiveness of KAM-based quantification of microscopic plastic strain.

Next, we measure TDR-based $\chi_{AC}$ for strained samples $S_{TS_1}^{0.05\%}$- $S_{TS_4}^{0.05\%}$ . The $S_{TS_1}^{0.05\%}$ sample closely resembles an unstrained sample, as strain effects are largely relaxed, and its magnetization response is comparable to that of an as-received (i.e., unstrained) $S^{0.05\%}$ steel sample. Fig. 4(a) shows that $\chi_{AC}$ increases with increasing strain and remains independent of *H* variation in the low-*H* regime (*H* is applied || to strain-direction in our experiments). This increase in $\chi_{AC}$ with strain is attributed to local lattice distortions (from KAM analysis). Strain anisotropy, induced by the magneto-elastic coupling in steel, changes the material's net magnetic anisotropy [60, 61]. Fig. 4 (b)(i) illustrates an unstrained sample ($\varepsilon = 0$) where magnetic anisotropy (MA) aligns differently from the direction of *H*. Corresponding *M*(*H*) curve for $\varepsilon = 0$ case is shown in Fig. 4(b)(iii) (black curve). When strain is applied ($\varepsilon \neq 0$), Fig. 4(b)(ii), strain-induced anisotropy reorients MA by an angle $\phi$ closer to the *H*-direction, which is the same as the applied strain-direction in our experiments. This, therefore, results in a steeper *M*(*H*) curve (red curve in Fig. 4(b)(iii) for *M*(*H*) measured along the MA direction), and exhibits enhanced $\chi$ for $\varepsilon \neq 0$. Fig. 4(c) confirms this, where our measured DC *M*(*H*) curves for the strained $S_{TS_1}^{0.05\%}$, $S_{TS_2}^{0.05\%}$ and $S_{TS_4}^{0.05\%}$ samples show an enhancing trend of $\chi$ with increasing strain.

The KAM analysis suggests strain-induced local distortions in the grain microstructure. However, magnetically this results in MA predominantly orienting in the strain-direction, resulting in the increasing $\chi$ of our artificially tensile-strained sample. Therefore, pure magneto-elastic strain effects in the 0.05% carbon steel sample alone cannot explain the decrease in $\chi$ observed in $S_{WQ}^{0.05\%}$ sample (Fig. 2(g) and Fig. S5(b) in SI-6). It may be worth commenting here on the increase in $\chi$ in our $S_{FA}^{0.05\%}$ sample compared to $S_{AR}^{0.05\%}$ (Fig. 2(f) and Fig. S5(a) in SI-6). The $S_{AR}^{0.05\%}$sample has dislocations due to the process of cold-rolling, which generate pinning sites obstructing DW motion while magnetizing the sample [2, 17, 62]. Slow annealing enhances crystallinity by enlarging grains and reducing dislocation density relative to the cold-rolled $S_{AR}^{0.05\%}$sample, and facilitates easy domain growth, resulting in higher $\chi$ for $S_{FA}^{0.05\%}$. Additionally, the $S_{AR}^{0.05\%}$sample exhibits a preferential alignment of grains along the [1 1 1] direction (Fig. 1(b), see the abundance of grains coloured in blue), which is ferrite's hard axis of magnetization [17]. Annealing aligns grains more uniformly, thereby boosting $\chi$, as observed in our experiments. We further explore below the decreased $\chi$ feature in $S_{WQ}^{0.05\%}$.

## 2.3. Evidence of vortex-like magnetic excitations in low-carbon steel

Motivated by the observation that quenching produces a mosaic of tiny and irregularly shaped crystallites with rapidly changing local magnetic anisotropy, we micromagnetically model such effects on steel's magnetic properties using MuMax3 [63], which solves the Landau-Lifshitz-Gilbert equation $\frac{d\vec{m}}{dt} = -\gamma\vec{m} \times \vec{H} + \alpha\,\vec{m} \times \frac{d\vec{m}}{dt}$ to give the evolution of magnetic moments (the damping constant $\alpha$ is taken as 0.5 in our simulations). An energy minimization procedure is employed in solving the differential equation, as detailed in references [64, 65]. A Voronoi tessellation algorithm generates grains of varying sizes representing polycrystalline samples (see SI-8 for discussion on micromagnetic modelling). From our IPF maps of low-carbon steels (Figs. 1(b)-(d)), we found a mosaic of crystallographically misoriented grains and thermal quenching of low-carbon steel significantly increases the extent of misorientation between the grains (Fig. 1(d)). Therefore, from the standpoint of magnetism, quenching produces a mosaic of rapidly changing local anisotropy direction from grain to grain. We therefore model the thermally treated low-carbon steels as possessing grains with different local MAs. In our simulation, we modelled two low-carbon steel samples (1024 nm × 1024 nm × 40 nm) with mean grain sizes of 40 nm and 10 nm ((Figs. 5(a)-(b)), to represent the $S_{FA}^{0.05\%}$ and $S_{WQ}^{0.05\%}$ samples respectively. The grain colours in Figs. 5(a) and (b) represent the varying MA direction of each grain (note that no periodic boundary condition was applied in order to take into account the realistic effect of demagnetization field in the finite-sized samples). For a more efficient computation of our structures, we scale down the size of the individual grains in our simulation compared to the actual average grain size in our samples, by a factor of $10^3$ while keeping the ratio of the grain-sizes between the $S_{FA}^{0.05\%}$ and $S_{WQ}^{0.05\%}$ approximately constant (recall, the mean grain sizes of the $S_{FA}^{0.05\%}$ and $S_{WQ}^{0.05\%}$ are 40 μm and 10 μm respectively, see table S4 in SI-3). We have found that our micromagnetic simulation results do depend on the size of the grains used within our simulations, as discussed next.

In our model, the MA of each grain is randomly chosen in the range (0.10 ± 0.01) MJ.m$^{-3}$ (compare cubic anisotropy of pure iron, $K_1$ ~ 0.1 MJ.m$^{-3}$ [66]). The exchange energy ($A$) is similarly considered constant within each grain (20 pJ.m$^{-1}$) [17] and allowed to vary from grain to grain randomly within a tolerance of 10% (i.e., 20 $\pm$ 2 pJ.m$^{-1}$). The saturation magnetization ($M_S$) used in the simulation is 400 kA.m$^{-1}$, obtained from experimental data and is assumed to be homogeneous throughout the geometry. Fig. 5(c) shows that the simulated $M(H)$ curve for the 40 nm grain-sized steel (i.e., Fig. 5(a)) aligns well with our experimentally measured $M(H)$ for the $S_{FA}^{0.05\%}$ sample, validating our micro-magnetic model. The simulated low-field $M(H)$ behaviour of 40 nm and 10 nm grain samples in Fig. 5(d) shows that the 40 nm grain-sized sample (i.e., 'annealed') has a higher $\chi$ than the 10 nm grain-sized (i.e., 'quenched') sample, with the susceptibility ratio consistent with our experimental observations from VSM and TDR.

Figs. 6(a)-(c) show the simulated field evolution of magnetic domains in the 10 nm grain-sized sample (which emulates the quenched sample). Also, see Fig. S9, SI-8 for the complete set and Fig. S8 for that of the 40 nm, or 'annealed' sample. In these images, colours represent the magnetization direction (note, the color wheel in the top-right corner, at each point of which, the magnetization direction is along the tangent to the circle), with white and black colours indicating moments along or opposite to $H$ respectively. As $H$ increases, magnetic moments align progressively with $H$, approaching saturation near 4.5 kOe (image appears whiter, Fig. S9 and S8). A closer examination of Figs. 6(a)-(c) reveals intriguing local features viz., persistent blackish spots at DW intersections around which we see a swirling of moments (encircled in Fig. 6(a)-(c)). Figures 6(a)→(b)→(c) show that as $H$ increases, the spot density decreases. With the vanishing of the spots, the domains expand (see the expansion of green-

yellowish and light blue regions in Figs. 6(b)→(c)). These spots do not move at low $H$ and DWs remain pinned at these sites. We also imaged the magnetic landscape of these low-carbon steels with magnetic force microscopy (MFM). The zero-field, dual-pass-phase image of the demagnetized state of $S_{AR}^{0.05\%}$, $S_{FA}^{0.05\%}$ and $S_{WQ}^{0.05\%}$ steel samples (Fig. 6(d)-(f)) show dark spots (magnetic features) of lateral size ~50 nm. These spots are comparable in size to the magnetic exchange length-scale of our low-carbon steel, $l_{ex} = \sqrt{2A/\mu_0 {M_s}^2}$ ~ 20 nm, where $A$ ~ 20 pJ.m$^{-1}$ and $M_S$ = 400 kA.m$^{-1}$. We find maximum spot density in $S_{WQ}^{0.05\%}$ (Fig. 6(f)), lower in $S_{FA}^{0.05\%}$ steel (Fig. 6(e)) and least in $S_{AR}^{0.05\%}$ (Fig. 6(d)). Fig. 6(g) shows a magnified view of these spots on $S_{WQ}^{0.05\%}$, which suggests a gradual and circularly symmetric swirl of magnetic moments: the central moments are oriented out-of-plane and the moments gradually orient along in-plane orientation towards the circumference. We also imaged the surface topography of the steel samples with atomic force microscopy (AFM) using a non-magnetic tip. AFM images (Fig. 6(h)-(j)) show uniform topographic features (r.m.s surface roughness of ~ 4 nm) and, more importantly, absence of any features similar to the dark spots observed in MFM. Therefore, we rule out the possibility of convolution of topographical surface features (hillocks/pits) in explaining the spots observed in MFM, thereby confirm their magnetic origin. Our simulations (Figs. 6(a)-(c)) reveal that dark localized spots, forming at DW intersections, display a swirling, non-collinear arrangement of magnetic moments with moment flipping at the centre (a quasi-singularity feature), resembling magnetic vortices [67]. Micro-magnetic energy calculations predict such swirling structures at the surface of soft magnetic materials (i.e., low anisotropy, so that DW width ~ $\sqrt{A/K}$ becomes broad and formation of swirls are favored) and estimate a cut-off of $3l_{ex}$ ~ 60 nm as radius of the swirl [68, 69]. The diameter of the dark spots in MFM (50-60 nm) are in good agreement with this estimate. MFM observations show the presence of such vortex-like magnetic excitations in all low carbon steel samples, though they are denser in thermally quenched samples. It is these vortex-like magnetic structures which pin DWs and restrict domain expansion with increasing $H$, consequently reducing the χ of quenched steel, as observed experimentally. Smaller grains (~10 nm) retain these vortex-like excitations up to higher fields (~3.5 kOe) than larger grains (~40 nm) (compare Fig. S9 and Fig. S8, respectively). Since such vortices form at DW intersections with varied moment directions, therefore they are more likely to form in quenched steel, with its small irregularly shaped grains and rapidly varying map of local MA distribution. We believe that thermal quenching produces massive HAGB fraction and consequently significant local strains in the $S_{WQ}^{0.05\%}$, which generate local lattice distortions and randomize local MA, thereby favouring formation of magnetic swirls, as have been observed in our MFM-based imaging of the magnetic landscape and micro-magnetic modelling.

## 3. Summary and Conclusions

Our study reveals that morphology, thermal treatments, and strain have a significant impact on the magnetic properties of low-carbon steel. Thermal annealing enlarges grains and enhances magnetic susceptibility, while quenching reduces both. Rapid quenching of low-carbon steel results in a significant increase in high-angle grain boundary fraction in the sample, generating a mosaic of small and irregularly shaped crystallographic grains with rapid spatial variation in randomly oriented magnetic anisotropy. Uniaxial mechanical strain increases magnetic susceptibility by altering magnetic anisotropy but does not explain the drop in susceptibility as observed for the quenched steel. Our micromagnetic modelling is motivated by the random variation in local magnetic anisotropy and shows localized regions of vortex-like swirling of magnetic moments near the domain edges. Magnetic force microscopy-based local magnetization images identify such magnetic excitations in quenched steel as domain wall pinning sites, microscopically causing the reduction in susceptibility. This work positions steel as a potentially useful material for magnetic device applications and also calls for further exploration of its vortex-like magnetic features.

## Acknowledgements

The authors acknowledge Centre for Nanoscience, IIT Kanpur for providing the MFM facility. SSB acknowledges Department of Science and Technology (DST)-SERB SUPRA program, the DST-AMT program of the Government of India, and IIT Kanpur for funding support. Suprotim Saha acknowledges the Prime Minister's Research Fellows (PMRF) scheme of the Ministry of Human Resource

Development, Government of India, for funding support. PCM thankfully acknowledges Mr. Sounavo Ghosh for fruitful discussions on electronic circuitry and simulations. PCM also thanks Mr. Bireshwar Roy, Mr. Prayas Sharma and Dr. Ankit Kumar for their involvements at different stages of the work.

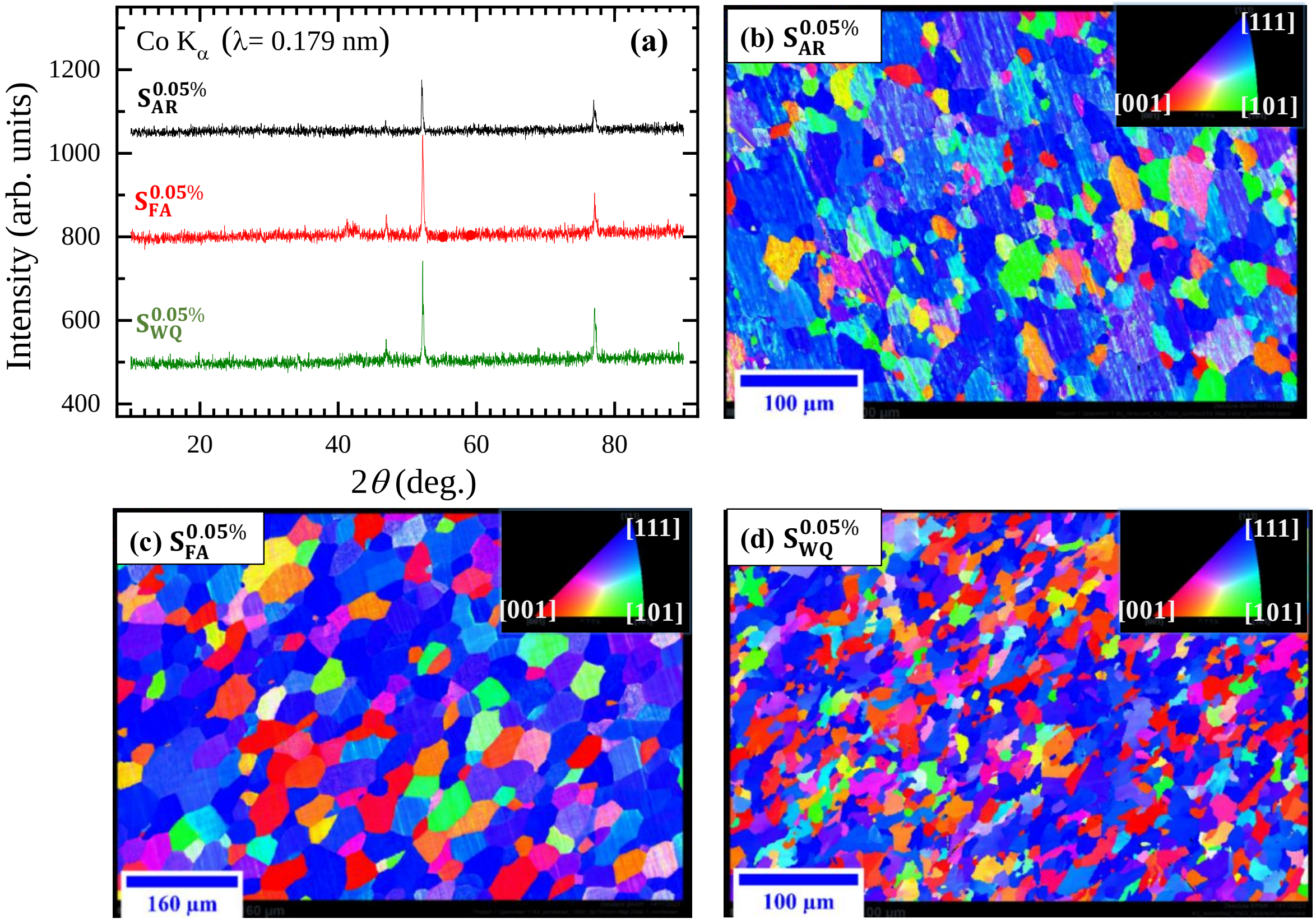

FIG 1. (a) X-ray Diffraction $2\theta$ scan on 0.05% C steel ($S^{0.05\%}$): as-received $S_{AR}^{0.05\%}$, furnace-annealed $S_{FA}^{0.05\%}$, and water quenched $S_{WQ}^{0.05\%}$ steel samples, with Co-K$_\alpha$ radiation. (b-d) IPF maps obtained from EBSD of $S_{AR}^{0.05\%}$ (b), $S_{FA}^{0.05\%}$ (c) and $S_{WQ}^{0.05\%}$ (d) steel samples. Note the color maps in (b-d) representing the crystallographic orientation of the grains along three cubic crystallographic axes. (cf. text for details)

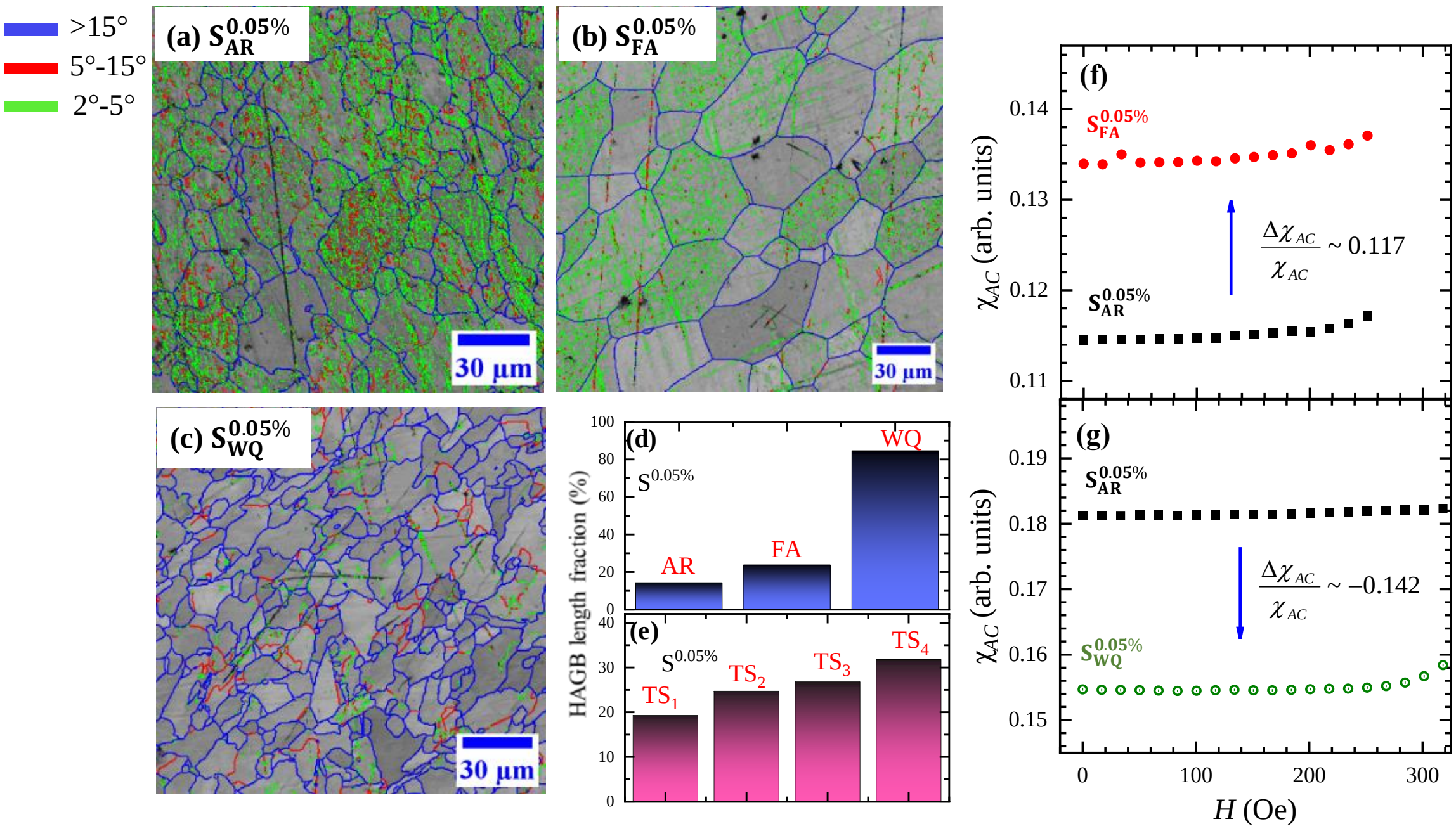

FIG 2. (a-c) Low and high-angle grain boundary maps in low-carbon steel samples: $S_{AR}^{0.05\%}$ (a), $S_{FA}^{0.05\%}$ (b), and $S_{WQ}^{0.05\%}$ (c). (color code for the grain-boundaries given in top-left corner). Note the abundance of high-angle grain boundary (HAGB) fraction (angle of grain-misalignment >15°, blue boundaries) in $S_{WQ}^{0.05\%}$, (c) compared to other two samples. (d) Comparison of HAGB length fractions in $S_{AR}^{0.05\%}$, $S_{FA}^{0.05\%}$ and $S_{WQ}^{0.05\%}$. (e) Comparison of HAGB length fractions in tensile-strained 0.05% carbon steel samples ($S_{TS1}^{0.05\%}$ - $S_{TS4}^{0.05\%}$). (f-g) Real part of AC susceptibility ($\chi_{AC}$) vs magnetic field ($H$) obtained from TDR-based measurement of $S_{FA}^{0.05\%}$ (f), and $S_{WQ}^{0.05\%}$ (g), steel samples, compared to $S_{AR}^{0.05\%}$ steel sample of equal dimensions. (cf. text for details)

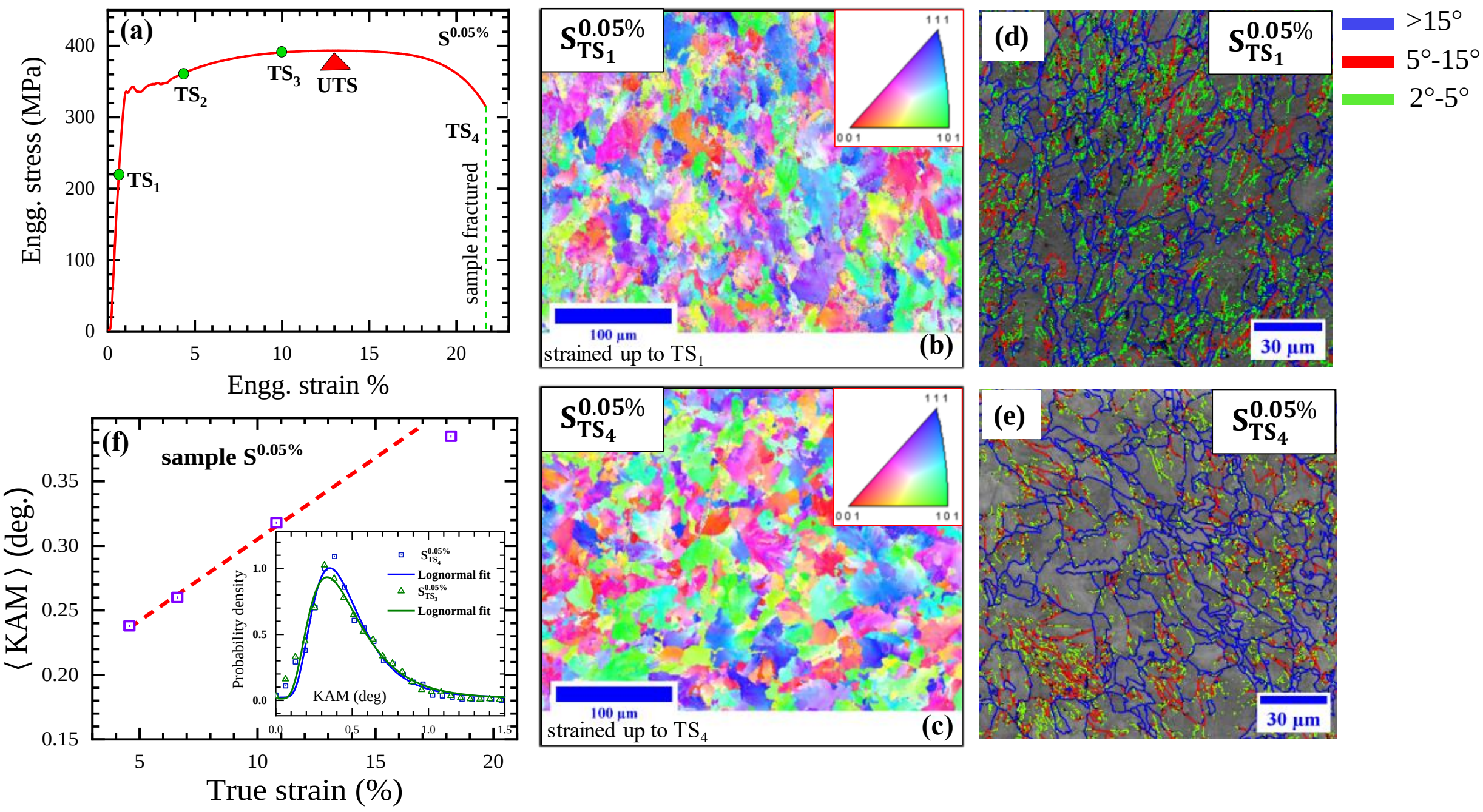

FIG 3. (a) Engineering stress (σ) vs engineering strain (ε) curve of a 0.05% carbon steel sample $S^{0.05\%}$. Green circles in the σ-ε plane labelled $TS_1$, $TS_2$, and $TS_3$ denote points up to which different $S^{0.05\%}$ steel samples have been tensile-strained. The fourth sample has been strained beyond Ultimate Tensile Strength (UTS) till fractured, see vertical drop (labelled $TS_4$). (b-c) IPF maps obtained from EBSD of strained steel samples, strained within elastic limit up to $TS_1$, $S^{0.05\%}_{TS_1}$ (b), and strained beyond UTS till fracture up to $TS_4$, $S^{0.05\%}_{TS_4}$ (c). Note the color maps in (b) and (c) representing the crystallographic orientation of the grains along three cubic crystallographic axes. (d-e) Low-and high-angle grain boundary maps in tensile-strained steel samples, strained up to $TS_1$, sample $S^{0.05\%}_{TS_1}$(d), and strained beyond UTS till fracture up to $TS_4$, sample $S^{0.05\%}_{TS_4}$ (e). Note the increase in HAGB fraction (blue lines) due to straining. (f) Mean misorientation angle (⟨KAM⟩) obtained from IPF maps plotted as a function of true strain %. (inset) Probability distribution of misorientation angles in two maximally strained samples, $S^{0.05\%}_{TS_3}$ and $S^{0.05\%}_{TS_4}$ with lognormal fits to data (cf. text for details).

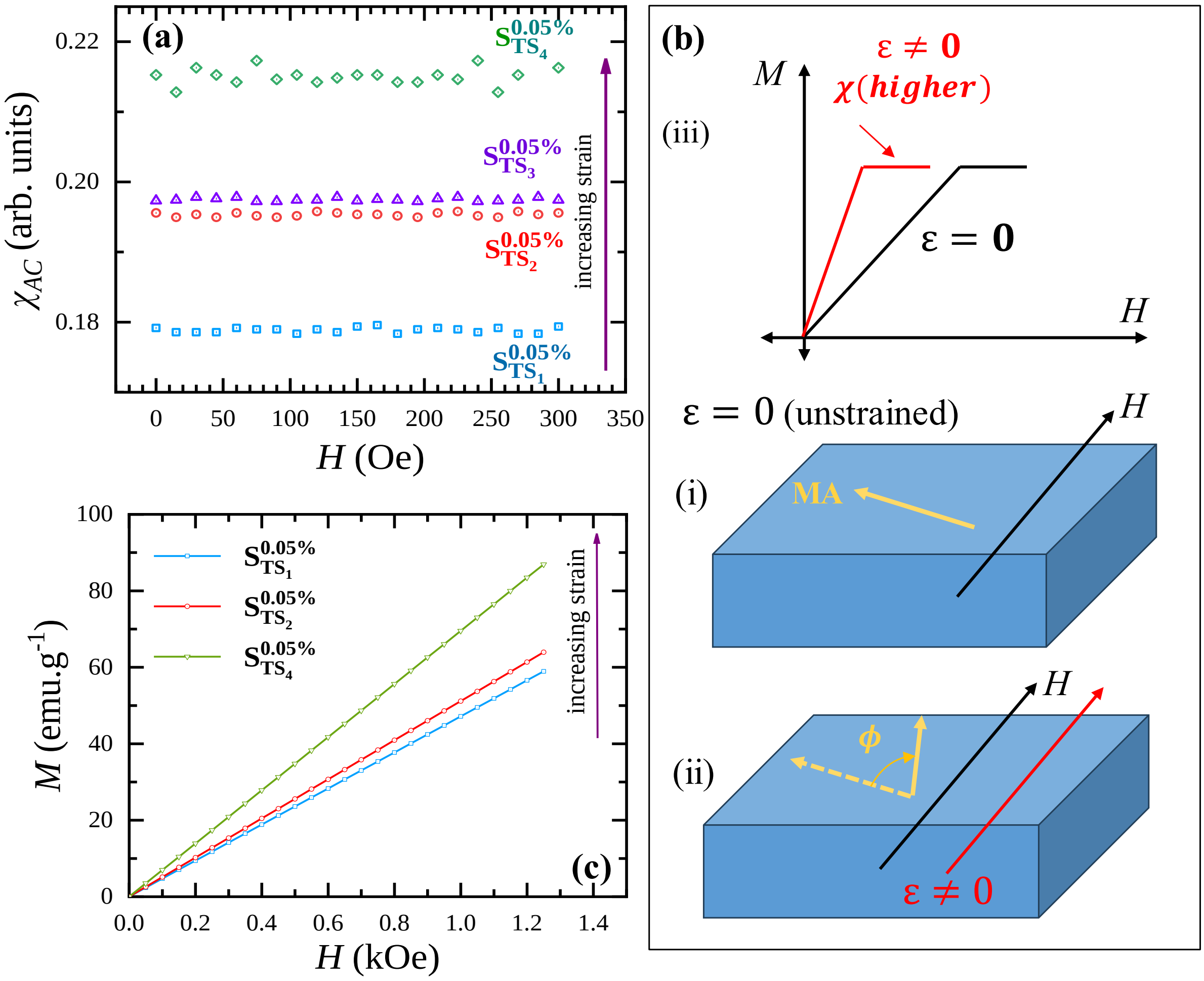

FIG 4. (a) $\chi_{AC}$ vs $H$ for steel samples with varying tensile-strain levels, $S^{0.05\%}_{TS_1}$ , $S^{0.05\%}_{TS_2}$ , $S^{0.05\%}_{TS_3}$ and $S^{0.05\%}_{TS_4}$. (b) Enhancement of susceptibility as a result of uniaxial tensile-strain (ε), unstrained case (i) and reorientation of anisotropy by angle $\phi$, solid yellow arrow in (ii) as a result of applied strain ε. Note that $H$ is applied along strain-direction. (c) Experimentally measured $M(H)$ curves for $S^{0.05\%}$ steel samples with varying tensile-strain levels, $S^{0.05\%}_{TS_1}$ , $S^{0.05\%}_{TS_2}$ and $S^{0.05\%}_{TS_4}$. Note that $H$ is applied along strain-direction in the experiments (cf. text for details).

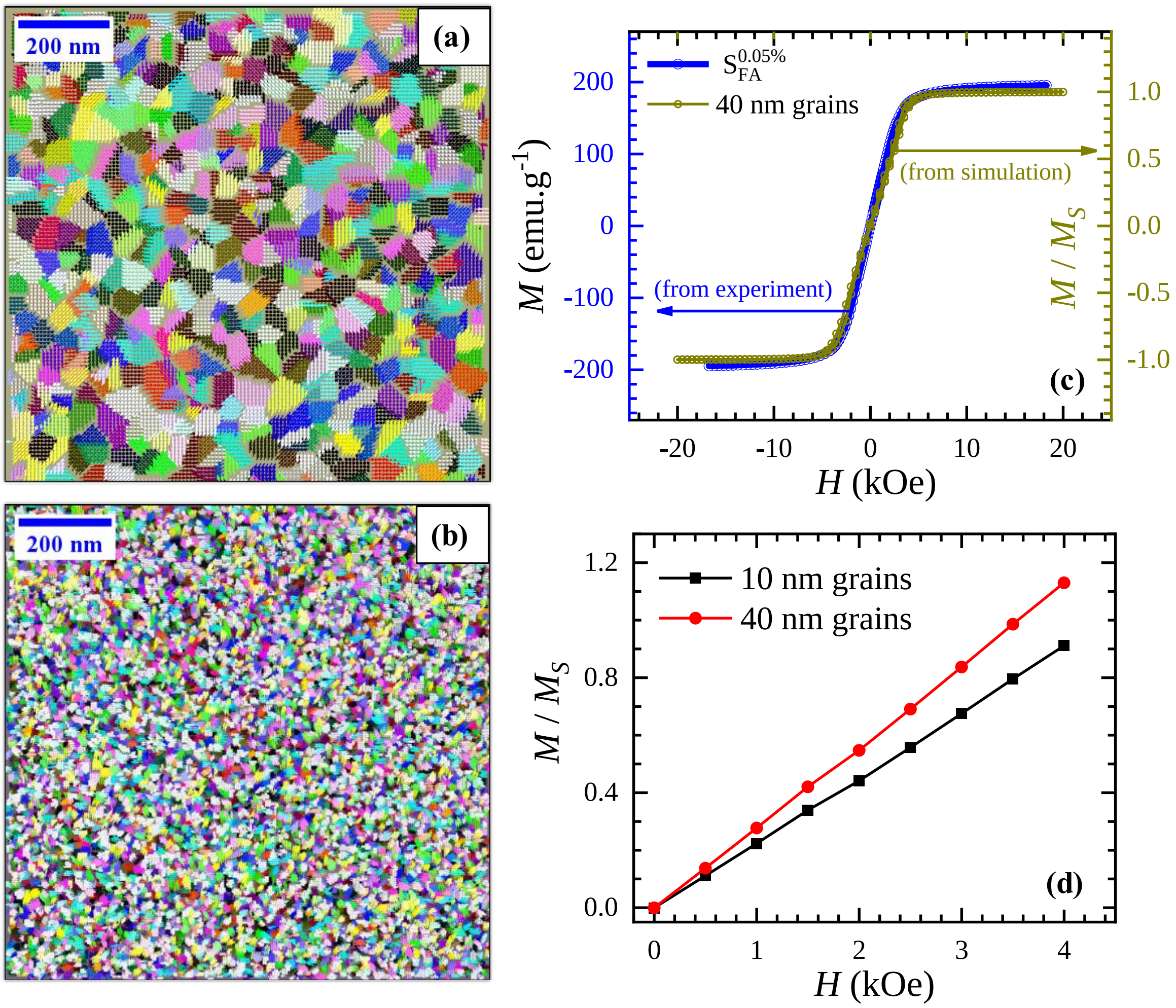

FIG 5. (a-b) Simulated microstructures of steel with varying sizes of grains, 40 nm (a), and 10 nm (b), with magnetic anisotropy values in the range (0.10 ± 0.01) MJ.m$^{-3}$ distributed randomly in each grain (denoted by different colors). (c) Comparison of experimentally obtained *M*(*H*) data for 0.05% carbon steel furnace-annealed sample ($S_{FA}^{0.05\%}$) with simulated *M*(*H*) result for 40 nm grain-sized steel sample i.e., Fig. 5(a). (d) Comparison of simulated *M*(*H*) results in the low-field regime for 40 nm and 10 nm grain-sized steel samples. (cf. text for details)

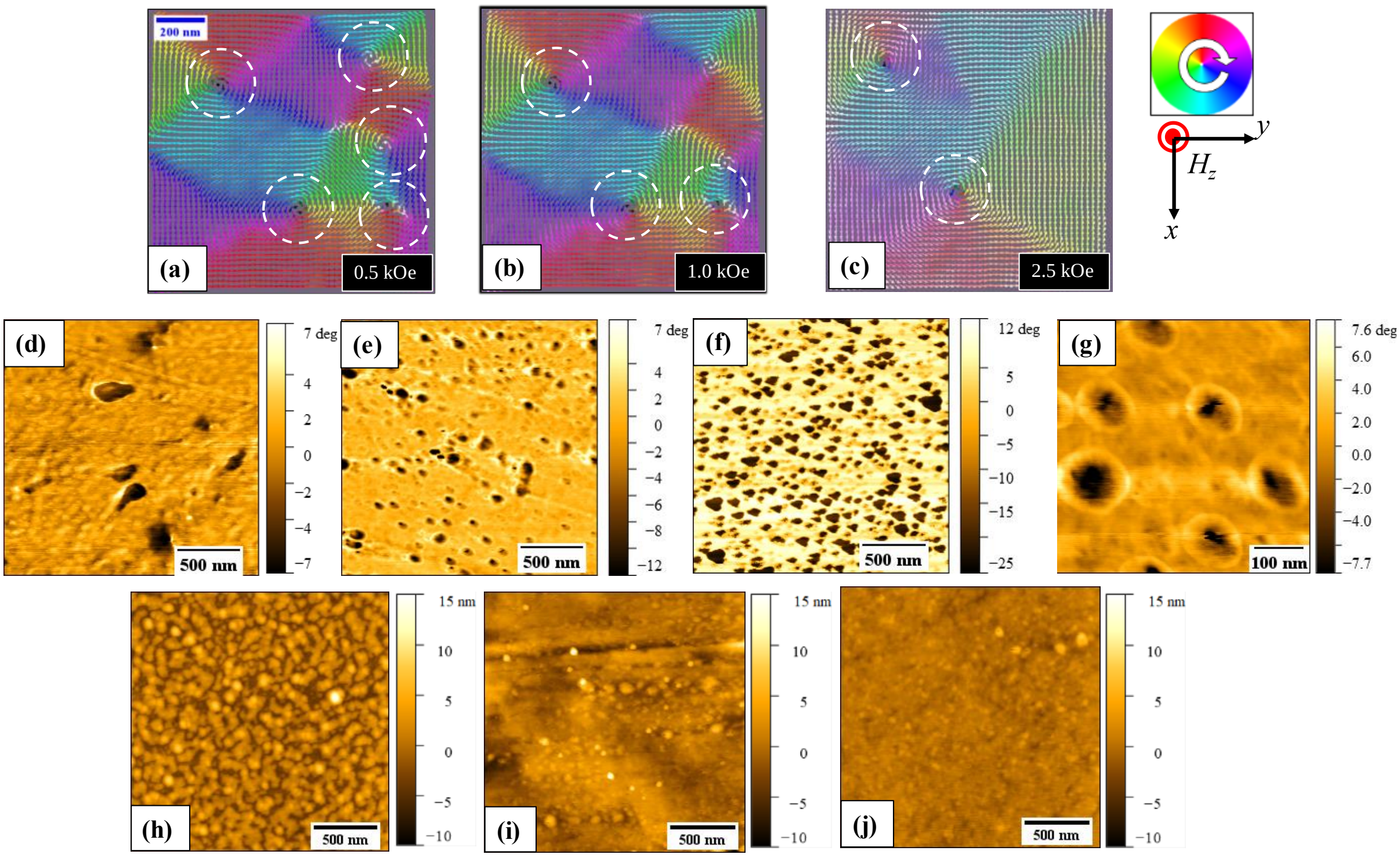

FIG 6. (a-c) Evolution of the simulated magnetic landscape ($\vec{M}$ (*x*, *y*)) as a function of field applied perpendicular to the plane ($H_z$). Magnetization direction is given by the color wheel in top-right corner, magnetization at each point on the color-wheel is tangential to the circle. White and black color denote magnetization along field-direction and antiparallel-to-field-direction, respectively. Note the localized dark spots (circled around in (a)-(c)) at the intersections of the domains. (d-f) Dual-pass-phase images in demagnetized state ($H$ = 0 Oe) obtained by MFM showing magnetic landscape in 0.05% carbon steel, as-received $S_{AR}^{0.05\%}$ (d), furnace-annealed, $S_{FA}^{0.05\%}$ (e), and water-quenched, $S_{WQ}^{0.05\%}$ (f) steel samples. Note the adjoining color bars in each image denoting change in phase of MFM in degree. (g) Magnified view of the magnetic vortex-like localized structures on $S_{WQ}^{0.05\%}$. Note the swirling of local moments from out-of-plane at the centre to in-plane at the circumference. (h-j) AFM maps showing surface topography on $S_{AR}^{0.05\%}$ (h), $S_{FA}^{0.05\%}$ (i), and $S_{WQ}^{0.05\%}$ (j) steel samples. Note the adjoining color bars in (h)-(j) denoting surface roughness (cf. text for details).

# Unravelling magnetic vortex-like excitations through rapid thermal quenching in low - carbon steel

P. C. Mahato[1], Suprotim Saha[1], Ritesh Kumar[1], D. Banik[2], K. Mondal[2], Shyam Kumar Choudhary [3], Ashish Garg[4], S. S. Banerjee[1,+]

[1]*Department of Physics, Indian Institute of Technology Kanpur, Kanpur, Uttar Pradesh 208016, India*

[2]*Department of Material Science and Engineering, Indian Institute of Technology Kanpur, Kanpur, Uttar Pradesh 208016, India*

[3]*Graphene Center, Tata Steel Limited, Jamshedpur, Jharkhand 831007, India.*

[4] *Department of Sustainable Energy Engineering, Kotak School of Sustainability, Indian Institute of Technology Kanpur, Kanpur, Uttar Pradesh 208016, India*

Corresponding authors email ID:

+: satyajit@iitk.ac.in

Supplementary Information – 1

Fig. S1(a) shows Fe-C equilibrium phase diagram with the relevant phases for this work viz., Austenite (γ), Ferrite (α), and Cementite marked. In steel, the C atoms are dissolved interstitially, which explains the higher carbon content in FCC austenite compared to BCC Ferrite. In our experiments, a steel sample with 0.05% carbon content ($S^{0.05\%}$, see on horizontal axis of Fig. S1(a)) was heated to 1000°C to Austenitize. Upon slow cooling, the austenite (FCC) phase transforms first into a mixture of ferrite and austenite (α + γ) as it crosses the upper-critical temperature line (marked $A_3$ in Fig. S1(a)). On further cooling, eventually the remaining austenite phase transforms into ferrite and cementite (orthorhombic, extremely low in quantity for our 0.05% carbon steel, hence not observed in our microstructural imaging,) as the temperature crosses the lower critical temperature line (marked $A_1$ in Fig. S1(a)).

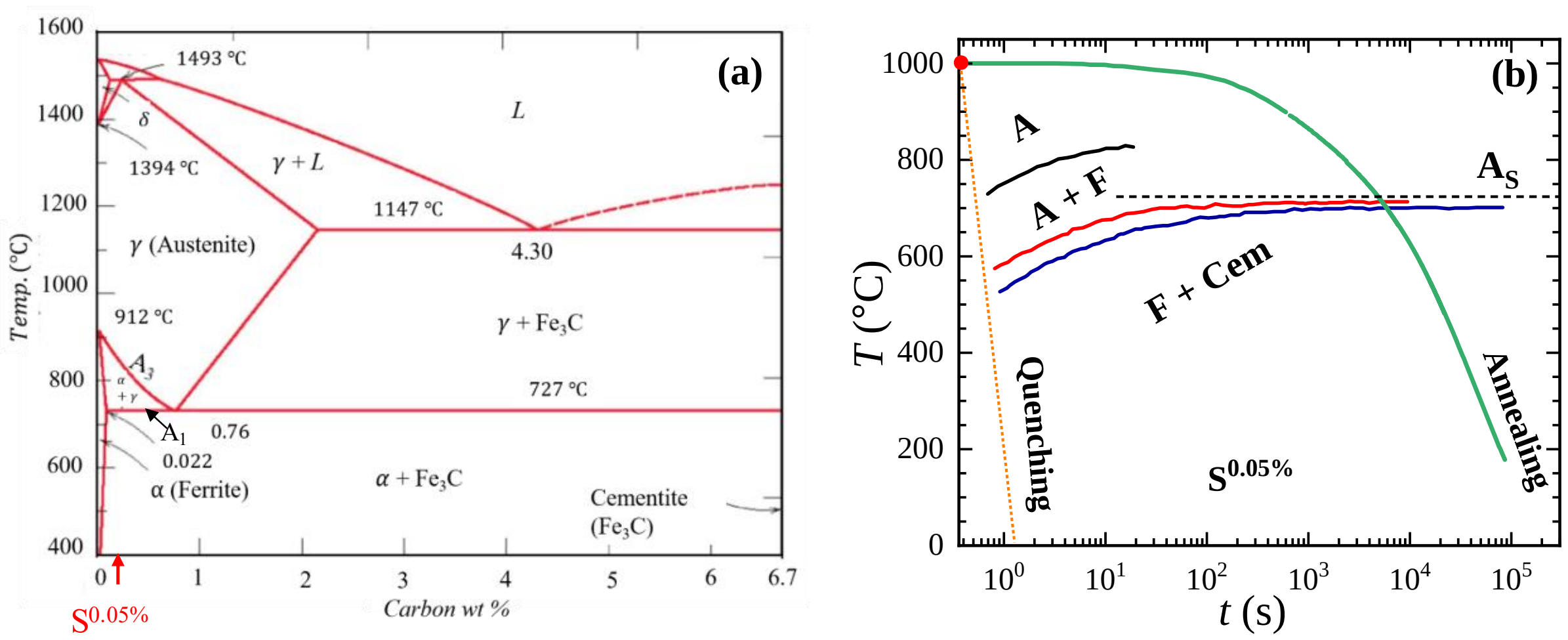

FIG S1 (a) Iron-carbon equilibrium phase diagram with relevant phases and phase-boundaries marked. (b) Time-temperature-transformation (TTT) diagram and continuous cooling transformation (CCT) curves for two different thermal processes, furnace annealing and water-quenching performed on a 0.05% carbon steel ($S^{0.05\%}$). Various phases e.g., Austenite, Ferrite and Cementite are represented by their initials A, F, and 'Cem' respectively. $A_S$ is the Austenite transformation temperature.

Optical Emission Spectroscopy (OES) was performed with a AMETEK-SPECTRO analyzer on steel samples to get the elemental composition.

Table S1. Elemental composition in weight % of $S^{0.05\%}$ steel

| C | Mn | Si | P | S | B | Cu | Ni | Cr | Nb | V | Mo | Ti | Al | Ce |
|---|---|---|---|---|---|---|---|---|---|---|---|---|---|---|
| 0.05 | 0.17 | 0.01 | 0.010 | 0.003 | 0.0001 | 0.005 | 0.019 | 0.018 | 0.001 | 0.001 | 0.001 | 0.001 | 0.057 | 0.08 |

## Supplementary Information –2

The as-received low carbon steel ($S_{AR}^{0.05\%}$) and thermally treated steel samples, viz., furnace-annealed ($S_{FA}^{0.05\%}$) and water-quenched samples ($S_{WQ}^{0.05\%}$) were metallographically prepared to view the microstructural details: ground with progressively fine grit SiC abrasive paper (600, 1000, 1500, 2000 and 2500-grits), polished with 5 μm alumina slurry and final polishing performed with virbometry (40 nm colloidal sillica) for ~ 8 hours to achieve a surface devoid of macroscopic scratches. These polished surfaces were imaged with Electron Backscattered Diffraction (EBSD) for viewing crystallographic and microstructural information. The same polished surfaces were eventually studied by magnetic force microscopy (MFM) and atomic force microscopy (AFM).

The polished surfaces achieved after vibromet polishing of the $S_{AR}^{0.05\%}$, $S_{FA}^{0.05\%}$ and $S_{WQ}^{0.05\%}$steel samples were chemically etched in order to image the microstructural details in optical micrography and scanning electron microscopy (SEM). All the $S^{0.05\%}$ steel samples were etched in 10% Nittal i.e., 10 % $HNO_3$ + 90 % methanol (by volume) for ~15 s (and eventually rinsing with deionized water). One usually discerns an optimally etched steel sample by noticing the onset of disappearance of the lustre of the initially shinny mirror-finished surface.

## Supplementary Information -3

X-ray Diffrcation (XRD) was performed on $S_{AR}^{0.05\%}$, $S_{FA}^{0.05\%}$ and $S_{WQ}^{0.05\%}$ steel samples of typical dimension 5 mm × 5 mm × 0.8 mm using a Panalytical X'Pert Pro MRD diffractometer with Co-Kα radiation (λ = 1.79 Å) at room temperature in the $2\theta$-range of (10º–90º) with a precision of 0.05º (see Fig. 1(a) in main manuscript). A word of caution: commonly used Cu-Kα radiation (λ = 1.5406 Å) should be avoided for characterizing steel samples (rather, samples containing Fe), as Fe atoms fluoresces in the X-rays produced by Cu target. We see that the peaks in the XRD spectra at $2\theta$ values of 46.9°, 52.3° and 77.1°, in the $\mathrm{S}_{\mathrm{AR}}^{0.05\%}$ are reproduced in our $\mathrm{S}_{\mathrm{WQ}}^{0.05\%}$ and $\mathrm{S}_{\mathrm{FA}}^{0.05\%}$ samples, suggesting the structural integrity (Ferrite BCC phase) of the steel samples is unmodified with furnace-annealing and water-quenching.

The polygonal shapes of the grains (Fig. S2(c) and (f)) and absence of peak-splitting in XRD (Fig. 1(a)) indicate absence of martensite phase in $\mathrm{S}_{\mathrm{WQ}}^{0.05\%}$ sample. The scratch-free polished surfaces of $\mathrm{S}^{0.05\%}$ steels were imaged in a Field-Emission Scanning Electron Microscope (JSM-7100F) with EBSD mode. The sample was tilted at 70° with respect to horizontal and raster-scanned with an electron beam (20 keV) at certain magnification (150X for $\mathrm{S}_{\mathrm{FA}}^{0.05\%}$, 250X for $\mathrm{S}_{\mathrm{WQ}}^{0.05\%}$ and $\mathrm{S}_{\mathrm{AR}}^{0.05\%}$sample). Diffraction of the backscattered electrons from the top 10-100 nm layer form Kikuchi patterns and these patterns, when fitted assuming Fe-bcc ($a = b = c = 2.87$ Å, space group Im$\bar{3}$m, space group no. 229) as a trial solution, yielded good match (see the pattern matching % in the following table-S3 for a typical annealed sample). Note that $\gamma$-Fe (fcc, $a = b = c = 3.66$ Å, space group Fm$\bar{3}$m, space group no. 225) structure was also tested as a trial solution, which resulted in poor matching (note the low phase fraction match and high Mean angular deviation, MAD for fcc), implying negligible Austenite content. Analyses of the EBSD images were performed in Atex [1].

EBSD of the tensile-strained steel samples $\mathrm{S}_{\mathrm{TS}_1}^{0.05\%}$- $\mathrm{S}_{\mathrm{TS}_4}^{0.05\%}$ (Fig. 3(b)-(c) and Fig. S7(a)-(d)) were performed in a Nova NanoSEM 450 FESEM. The Kikuchi patterns were fitted with Ferrite as trial solution, as before.

Table S3: Phase fraction from EBSD for furnace-annealed sample

| Phase Name | Phase Fraction (%) | Phase Count | Min Band Contrast | Max Band Contrast | Mean MAD | Standard Deviation MAD | Min MAD | Max MAD |
|---|---|---|---|---|---|---|---|---|
| Iron bcc | 98.43 | 636879 | 31.00 | 183.00 | 0.58 | 0.14 | 0.14 | 2.00 |
| Iron fcc | 0.00 | 11 | 68.00 | 122.00 | 0.87 | 0.27 | 0.43 | 1.34 |
| Zero Solutions | 1.57 | 10155 | 0.00 | 237.00 | | | | |

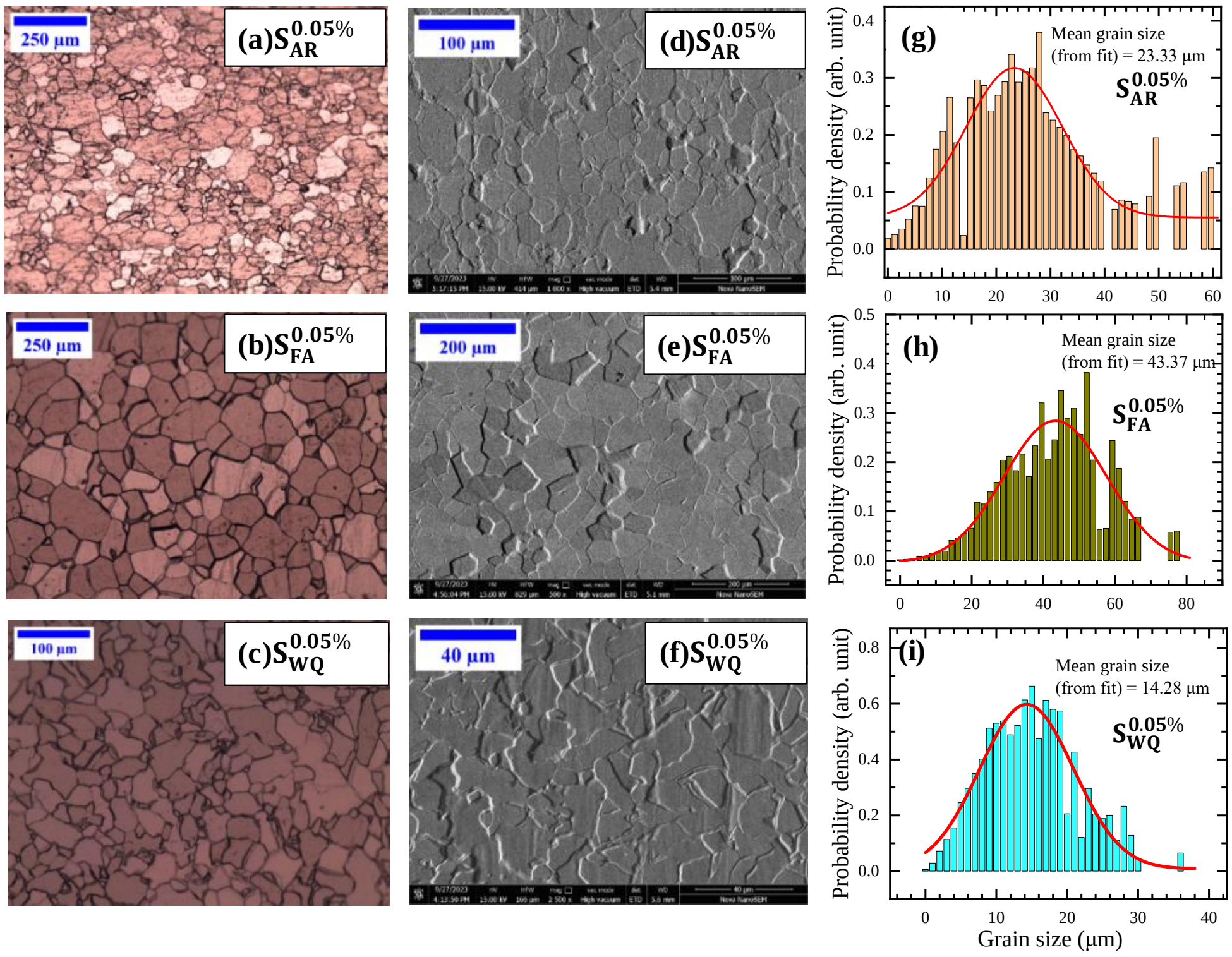

FIG S2 (a)-(c) Optical micrographs of the grains in 0.05% carbon steel ($S^{0.05\%}$) as-received $S_{AR}^{0.05\%}$ (a), furnace-annealed $S_{FA}^{0.05\%}$ (b), and water-quenched $S_{WQ}^{0.05\%}$ (c) steel samples. (d-f) SEM images of the grains in $S_{AR}^{0.05\%}$ (d), $S_{FA}^{0.05\%}$ (e), and $S_{WQ}^{0.05\%}$ (f) steel samples. (g-i) Grain-size statistics obtained from respective IPF maps along with Gaussian fits to data. Note the mean grain-size given in each image.

Table S4: Comparison of grain-sizes, high and low angle grain boundaries, DC (VSM, $\chi$) and AC (TDR, $\chi_{AC}$) susceptibilities in the $S_{AR}^{0.05\%}$ and the two kinds of thermally treated steels ($S_{FA}^{0.05\%}$ and $S_{WQ}^{0.05\%}$)

| | Mean grain size (μm), Figs. 1 (b)-(d) | HAGB fraction (>15°), Figs. 2(a-c) | LAGB fraction (2° - 15°), Figs. 2(a-c) | % change in χ (compared to $S_{AR}^{0.05\%}$), Figs. S5 (a)-(b) | % change in $\chi_{AC}$ (compared to $S_{AR}^{0.05\%}$), Figs. 2 (f)-(g) |
|---|---|---|---|---|---|
| $S_{AR}^{0.05\%}$ | $23.33 \pm 0.77$ | 14.3 | 85.7 | 0 | 0 |
| $S_{FA}^{0.05\%}$ | $43.37 \pm 0.92$ | 23.8 | 76.2 | 11.7 | 11.7 |
| $S_{WQ}^{0.05\%}$ | $14.38 \pm 0.34$ | 84.6 | 15.4 | -14.9 | -14.2 |

The vibromet-polished scratch free surfaces were metallographically etched and imaged with a SEM (Nova NanoSEM) and an optical microscope (Olympus BX51M) to reveal the microstructural details. Fig. S2 shows a compilation of microstructural images on different steels: $S_{AR}^{0.05\%}$(Fig. S2(a), (d)), $S_{FA}^{0.05\%}$ (Fig. S2(b), (e)), and $S_{WQ}^{0.05\%}$ (Fig. S2(c), (f)). Fig. S2 (g-i) represents a statistical distribution of the grain-sizes in $S_{AR}^{0.05\%}$, $S_{FA}^{0.05\%}$, and $S_{WQ}^{0.05\%}$ steel respectively, obtained from IPF maps of respective samples. Note the clear increase in grain-sizes in the furnace-annealed sample $S_{FA}^{0.05\%}$ and shrinking of grains in water-quenched steel $S_{WQ}^{0.05\%}$, in comparison with as-received $S_{AR}^{0.05\%}$steel (see Table S4 for quantification).

.

Supplementary Information -4

The Vickers hardness test was performed on $S_{AR}^{0.05\%}$, $S_{FA}^{0.05\%}$ and $S_{WQ}^{0.05\%}$ steel samples in a micro-Vickers hardness tester (Banbros Engg., model: VHS-1000 AT) with a load of 100 gm-force. Fig. S3(a)-(c) shows the indents as results of hardness-testing and Table S5 compiles the measured hardness for these 3 samples. Note the significant increase in hardness (~100%) for $S_{WQ}^{0.05\%}$ (or, coarse-grained) steel.

Table S5: Comparison of hardness of $S_{AR}^{0.05\%}$and thermally treated steels ($S_{FA}^{0.05\%}$ and $S_{WQ}^{0.05\%}$)

| $S_{AR}^{0.05\%}$ (HV) | $S_{FA}^{0.05\%}$ (HV) | $S_{WQ}^{0.05\%}$ (HV) |
|---|---|---|
| 100.6 | 90.1 | 202.5 |

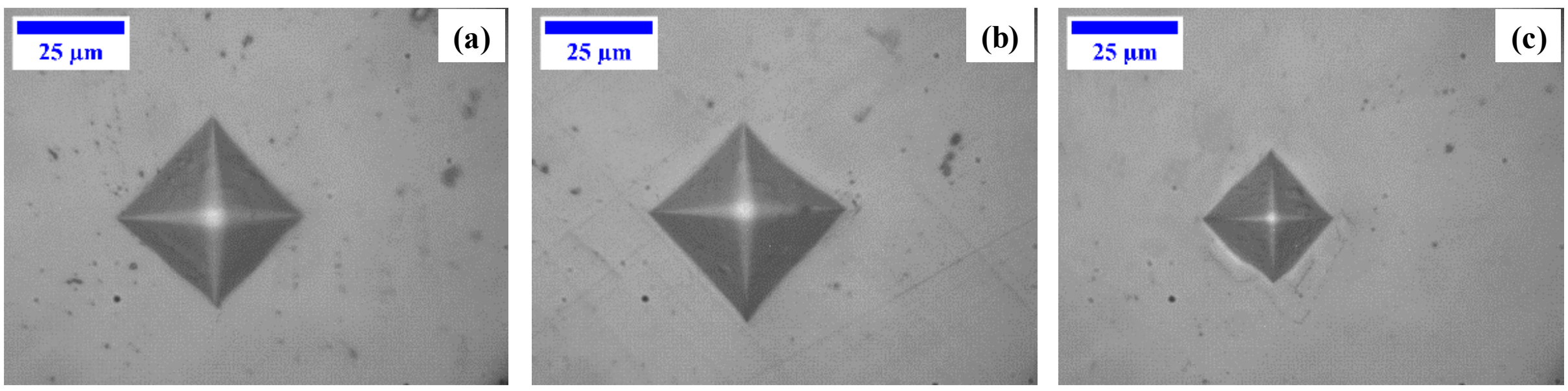

FIG S3 (a)-(c) Indentations made for Vickers hardness test with a 100 gram-force load on a 0.05% carbon steel ($S^{0.05\%}$): as-received $S_{AR}^{0.05\%}$ (a), furnace-annealed $S_{FA}^{0.05\%}$ (b), and water-quenched $S_{WQ}^{0.05\%}$ (c) steel samples.

Supplementary Information -5

Tunnel diode resonator (TDR) based ac susceptibility ($\chi_{AC}$) measuring technique uses a tunnel diode in conjunction with a simple tank ($LC$) circuit. The Negative Differential Resistance (NDR) property of the Tunnel Diode compensates for the resistive losses in the tank circuit, thereby sustaining the resonant oscillations ($f_0 \sim 1/2\pi\sqrt{LC}$). Tunnel diodes are heavily doped p-n junction diodes with extremely thin depletion region (~ nm) which makes quantum mechanical tunnelling of electrons possible from degenerately doped n-side to p-side. As a result, at low forward bias, a tunnelling current is observed, as opposed to ordinary p-n junction diode [2]. With proper biasing and optimization of circuit parameters, TDR can produce steady and sustained oscillation in the radio frequency (RF) band. For sustained oscillations to exist, the circuit parameters have to be chosen in such a way that the average power per cycle supplied by the diode is equal to the power dissipated in the rest of the circuit [3].

A magnetic sample (e.g., steel), when placed inside the tank inductor coil ($L$), experiences a small time($t$)-varying magnetic field ($H(t)$) generated due to the current in the surrounding solenoid. This produces a time-varying magnetization ($M(t)$) in the sample, which, in the most general case, is not in phase with $H(t)$, resulting in a real ($\chi_{Re}$) and an imaginary component ($\chi_{Im}$) of magnetic susceptibility (i.e. $\chi_{AC} = \chi_{Re} + i\chi_{Im}$). A magnetic sample inserted into the solenoid causes the effective magnetic flux and inductance $L$ to change which in turn causes a shift in $f_0$. If the change in inductance ($\Delta L/L$ << 1) is small, it can be shown $\frac{\Delta f_m}{f_0} \approx -\frac{1}{2}F\chi_{Re}$ where $\Delta f_m$ is the shift in resonant frequency due to sample and $F$ is the filling factor, defined as the ratio of the sample-volume ($V_S$) and the inductor coil volume ($V_C$). Therefore, measuring $\Delta f_m$ gives an estimate of $\chi_{AC}$ of the magnetic specimen. TDR-based technique has been extensively used in measurement of magnetic susceptibility of crystalline samples and magnetic thin films [4, 5], quantification of London penetration depth and critical field in variety of superconductors [6, 7], observation of Shubnikov-de Haas oscillation [6] and de Haas-van Alphen oscillations [8, 9] of purely quantum origin.

A Ge-based tunnel diode 1N3712 has been used in the present work, whose dc current($I$)- voltage ($V$) parameters are summarized in table S6 (see, e.g., ref. [2] for the significance of the parameters).

Table S6: Characteristic current ($I$) and voltage ($V$) parameters of Ge-based tunnel diode 1N3712

| Peak Current, $I_P$ (mA) | Peak Voltage, $V_P$ (mV) | Valley current, $I_V$ (mA) | Valley voltage, $V_V$ (mV) |
|---|---|---|---|
| $1.01 \pm 0.02$ | $50 \pm 2$ | $0.20 \pm 0.01$ | $300 \pm 7$ |

The circuit schematic of the TDR-based susceptometer, following Clover et al. [3] is shown in Fig. S4(a), which has been simulated in PLECS [10]. We have fabricated the circuit on a single layer printed circuit board (PCB) (see table S7 for typical values of circuit parameters). The inductor ($L$) was constructed by winding 12 turns of a copper wire (SWG 23) on a hollow cylinder (made of delrin) of outer diameter 3.3 mm (see Ref. [11] for the understanding the role of each circuit component). Fig. S4(d) shows the bias voltage dependent frequency output which is in reasonable agreement with the simulation results. All experiments were done at keeping the source voltage fixed at 0.615 V, where the frequency reaches a flat maximum (see Fig. S4(d)) and therefore, is impervious to small fluctuations in bias. The peak tunnelling current of 1N3712 is ~ 1 mA (Table S6), which is the maximum current through $L$, producing a magnetic field of ~ mOe. This is, therefore, the typical field experienced by the sample when placed inside the tank coil.

Table S7: Optimized circuit parameters for steady and sustained oscillations from TDR (see Fig. S4(a))

| $R_1$ (Ω) | $R_2$ (Ω) | $C_1$ (pF) | $C_3$ (nF) | $C_4$ (nF) | $L_C$ (μH) | R (Ω) | L (μH) | C (pF) |
|---|---|---|---|---|---|---|---|---|
| 330 | 330 | 10 | 10 | 8 | 122 | 100 | 0.75 | 5 |

We have calculated the net impedance $Z(f)$ of the circuit (Fig. S4(b-c)) and solved for resonant frequency which comes very close to the experimentally obtained values (~72 MHz at 0.615 V, compare to the peak-position in Fig. S4(b)), (the diode is simply replaced by a negative resistor of ~ 500 Ω i.e., $dI/dV \sim 2 \times 10^{-3}$ mS in the analytical expression). Note that the phase remains constant at the region of operation (Fig. S4(c)).

As a hallmark of performance and sensitivity of our TDR suscpetometer, we study a superconducting transition in $Bi_2Sr_2Ca_2Cu_3O_{10}$ (BSCCO 2223) polycrystalline sample. Fig. S4(e) shows the $f$ vs $T$ warming data in zero field ($T$ is measured with a pre-calibrated CERNOX resistive sensor). The entrance into superconducting state is characterized by a large and rapid increase in $f$ ($\Delta f_m$ = 6.348 MHz.g$^{-1}$) at 94.90 K. In the superconducting state, the material expels any magnetic flux from entering into its interior (Meissner screening) barring a thin surface layer called penetration depth ($\lambda_L$) [12]. This results in significant decrease in effective flux area (and therefore, $L$), which is reflected as a rapid increase in $f$. As $T$ approaches the superconducting transition temperature ($T_C$), the Meissner screening gets progressively weaker and $f$ gradually drops to the $T$-independent background value, which is practically the frequency corresponding to the empty solenoid. We corroborate this result with magnetization ($M$) versus $T$ data measured for the same sample in a SQUID magnetometer (CRYOGENIC, UK) (see inset of Fig. S4(e)), which shows a sharp drop in $M$ starting at 110 K. $T_C$ is marked by the peak in $dM/dT$ at 96.3 K, in close agreement with our TDR results. This unambiguously establishes the efficiency of our TDO sensor as a high sensitivity susceptometer.

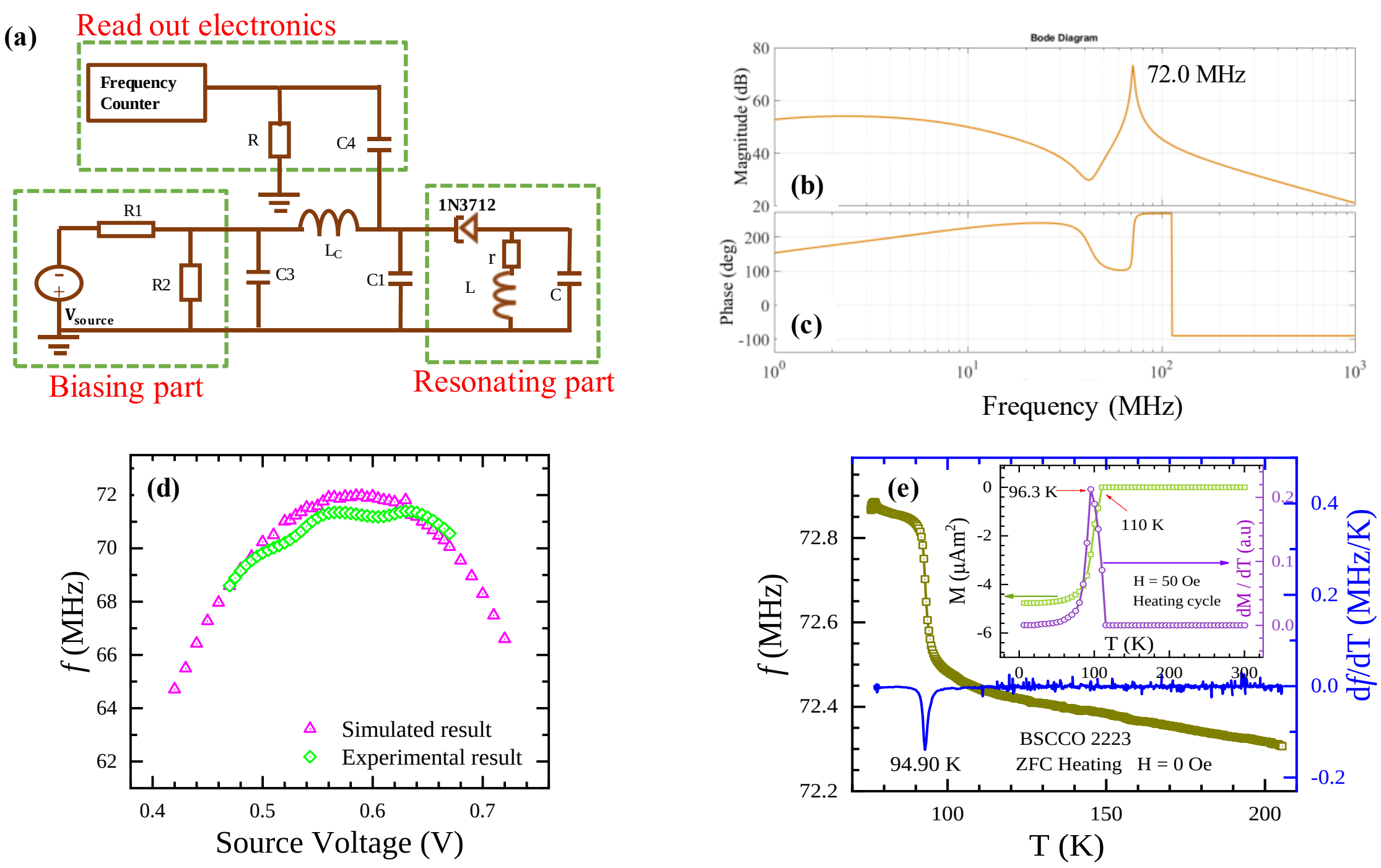

FIG. S4 (a) TDR complete circuit, three basic modules viz., biasing part, resonating part and read out electronics highlighted (note the values of components in table-S7). (b-c) Magnitude (b), and phase (c), of complex impedance ($Z$) of TDR circuit as a function of frequency ($Z(f)$). The tunnel diode is represented by a negative resistor -500 Ω in the analytical expression of $Z(f)$, rest circuit parameters are same as shown in table S7. (d) Comparison between experimentally obtained frequency vs voltage results with simulation. (e) A superconducting transition in polycrystalline BSCCO 2223 observed with TDR, the transition appears as a large and rapid increase in $f$ (see the clear notch in $df/dT$ at 94.9 K, blue curve, right axis). (inset) $M$ vs $T$ corresponding to the same superconducting transition observed with a SQUID magnetometer showing a sharp peak in $dM/dT$ at 96.3 K. (cf. text for details)

Supplementary Information- 6

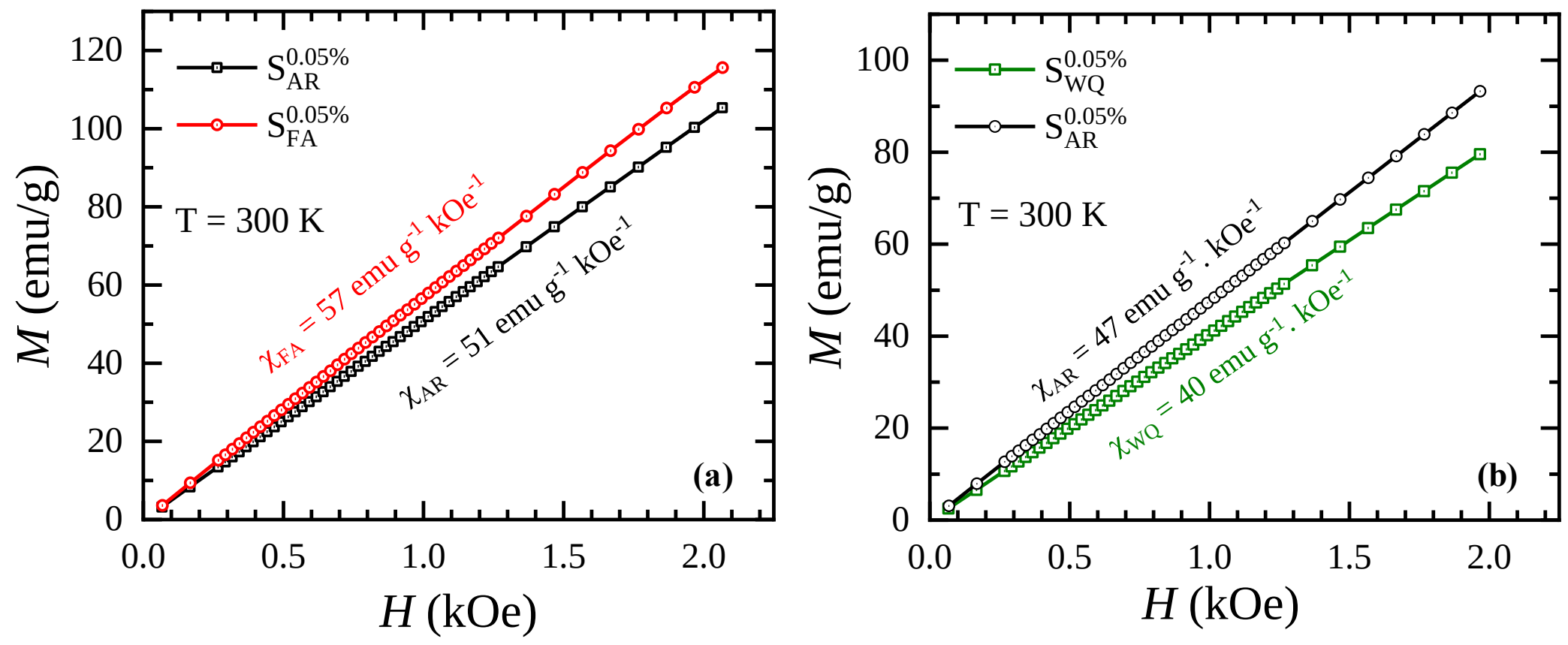

FIG S5. (a-b) *M*(*H*) results obtained from VSM comparing dc magnetic susceptibility of furnace-annealed 0.05% C steel, $S_{FA}^{0.05\%}$ (a), and water-quenched 0.05% C steel $S_{WQ}^{0.05\%}$ (b) samples with as-received steel sample $S_{AR}^{0.05\%}$ of equal dimensions. Note the dc susceptibility values for each *M*(*H*) curve are given in figures.

DC magnetization (*M*) of all steel samples were measured as a function of applied magnetic field (*H*) in a commercial VSM (ADE technologies, model EV7). Figs. S5 (a)-(b) show the first-quadrant *M*(*H*) behaviors of $S_{FA}^{0.05\%}$ and $S_{WQ}^{0.05\%}$ steel samples respectively, in comparison with $S_{AR}^{0.05\%}$steel sample of equal dimension. Note that we cut all samples to identical geometric dimensions for these *M*(*H*) measurements (5 mm × 5 mm × 0.8 mm) to keep the demagnetization factor identical. For all measurements, we initially recorded the *M*(*H*) of an as-received $S_{AR}^{0.05\%}$ sample and after undergoing thermal treatment, we measured the *M*(*H*) of the same sample again. Susceptibility (defined as $\chi = \frac{dM}{dH}$), from the DC magnetization *M*(*H*) measurements show an important correlation with grain-sizes: the $S_{FA}^{0.05\%}$sample with larger grain size has a higher $\chi$ compared to $S_{AR}^{0.05\%}$steel, while the $S_{WQ}^{0.05\%}$ sample with smaller grains has a lower $\chi$ compared to the $S_{AR}^{0.05\%}$ sample (see table S4, SI-3).

Supplementary Information- 7

In order to measure the stress-strain curve, $S^{0.05\%}$ steel samples were cut and patterned with electrical discharge machining (dimensions given in Fig. S6 below), following two relevant normative references ISO 6892-1:2016 'Metallic materials - Tensile testing, Part 1: Method of test at room temperature" and ASTM

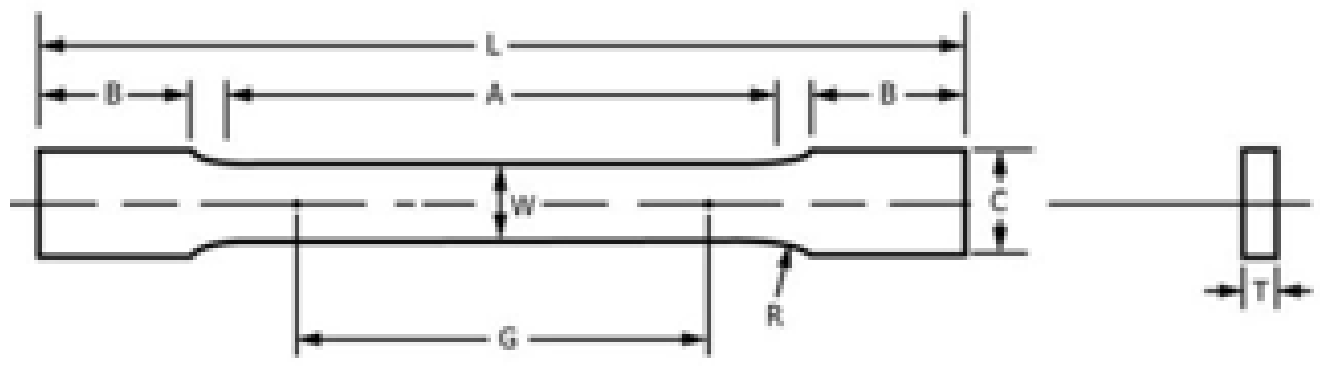

G (gauge length) = 25.00 mm
W (width) = 6.0 mm
T (thickness) = 1.91 mm
A = 32 mm
B = 30 mm
C = 10 mm
L = 100 mm
R (radius of fillet) = 6 mm

FIG S6 Dimensions of a 0.05% carbon steel $S^{0.05\%}$ work piece used for tensile straining

E8/E8M-16a ''Standard Test Methods for Tension Testing of Metallic Materials" [13]. The strain rate was kept 0.001 $s^{-1}$. For EBSD, ferrite (bcc) and cementite (orthorhombic, $a$ = 5.0837 nm, $b$ = 6.7475 nm and $c$ = 4.5165 nm, orthorhombic, space group *Pnma* [14]) were used as trial solution to fit the Kikuchi bands. Figs. S7 (a-d) show the Inverse Pole Figure (IPF) maps for samples $S_{TS_1}^{0.05\%}$ - $S_{TS_4}^{0.05\%}$ (as per the labelling scheme detailed in main text, see Fig. 3(a) and relevant discussions therein). These images were analyzed to calculate the local Kernel Average Misorientation (KAM) defined as $KAM\ (P) = \frac{1}{4}\sum_{i=1}^{4}\theta\ (P, P_i)$ where $\theta(P, P_i)$ is the crystallographic misorientation between any point $P$ and its neighboring points $P_i$ (there are 4, as we considered square grids) [15]. Note that a grain tolerance angle of 5° is used in the algorithm to exclude misorientations associated with grain-boundaries. In other words, only the grids having a misorientation less than this tolerance limit with their neighboring 4 grids are considered. Inset of Fig. 3(f) shows the probability distribution of misorientation angles in two maximally strained samples: strained up to $TS_3$ and strained > UTS i.e., $TS_4$ (vide Fig. 3(a) in manuscript). The probability distribution curves are fit with log-normal distribution. The mean KAM obtained from these fits are plotted as a function of true macroscopic strain (Fig. 3(f) in manuscript). Figs. S7(e-h) show the low and high-angle grain boundary maps for the tensile-strained steel samples $S_{TS_1}^{0.05\%}$ - $S_{TS_4}^{0.05\%}$.

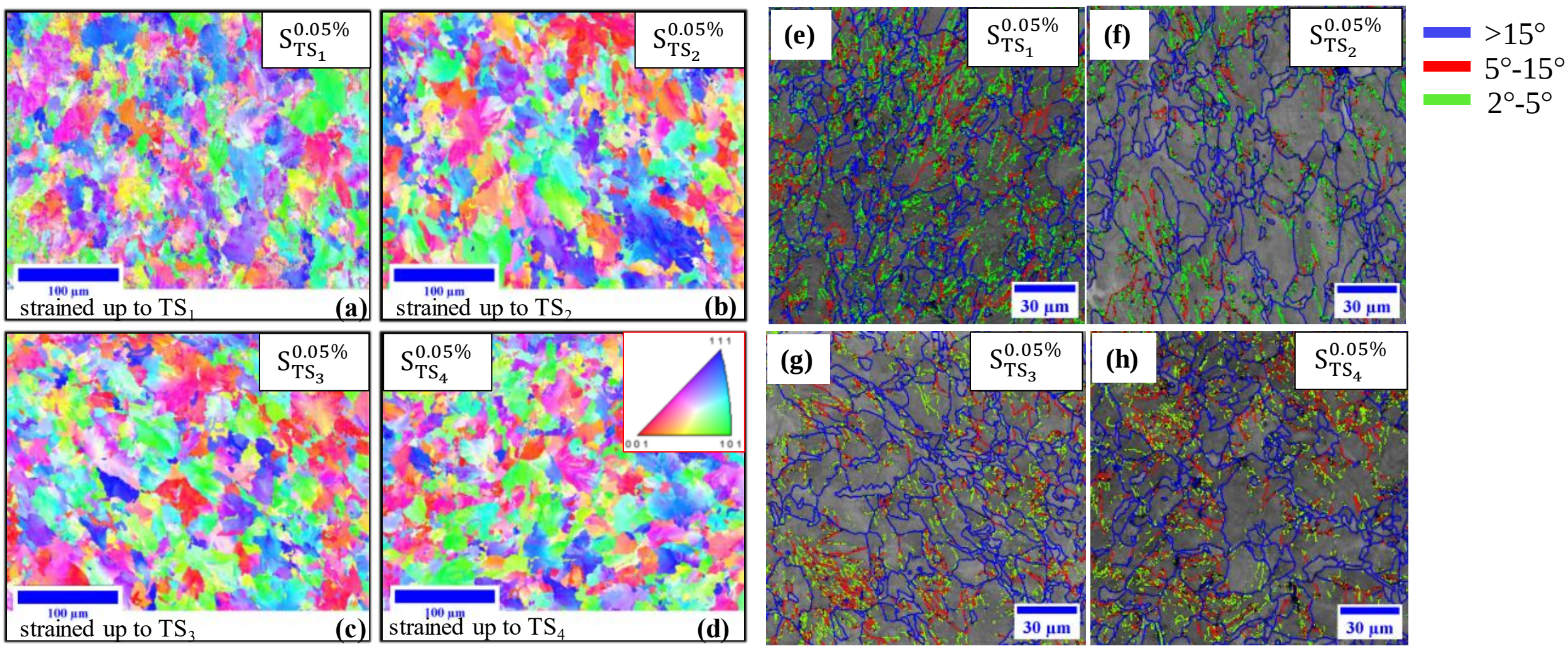

FIG S7. (a)-(d) IPF maps obtained from EBSD of tensile-strained steel samples, strained up to $TS_1$, sample $S_{TS_1}^{0.05\%}$ (a), strained beyond elastic regime up to $TS_2$, sample $S_{TS_2}^{0.05\%}$ (b), strained deep in the plastic regime near UTS, up to $TS_3$, sample $S_{TS_3}^{0.05\%}$(c) and strained beyond UTS till fracture up to $TS_4$, sample $S_{TS_4}^{0.05\%}$ (d) (cf. text and Fig. 3(a) for details). Note the color map in (d) denoting three cubic crystallographic axes for (a-d). (e)-(h) Grain boundary maps for four tensile-strained samples: $S_{TS_1}^{0.05\%}$ (e), $S_{TS_2}^{0.05\%}$ (f), $S_{TS_3}^{0.05\%}$ (g) and $S_{TS_4}^{0.05\%}$ (h), note the color map for grain-boundaries in the top-right corner for (e-h).

## Supplementary Information – 8

We have attempted to visualize the evolution of magnetic domains in $S^{0.05\%}$ steels consisting of varying grain sizes by means of a micro-magnetic framework. The micro-magnetic modelling were performed in a GPU-accelerated simulation-package Mumax3 [16]. Magnetization is treated as a classical vector ($\vec{m}$) and the basic energy functional ($E$) required to describe the magnetization process contains the exchange, magnetostatic, anisotropy and Zeeman terms. The gradient of the energy functional with respect to magnetization gives the effective field ($\vec{H}_{eff} \propto \nabla_{\vec{m}} E$ ) which then governs the magnetization dynamics through the Landau-Lifschitz-Gibert equation: $\frac{d\vec{m}}{dt} = -\gamma\vec{m} \times \vec{H}_{eff} + \alpha\,\vec{m} \times \frac{d\vec{m}}{dt}$

The damping constant $\alpha$ is taken as 0.5 in our simulations. An energy minimization procedure is employed in solving the differential equation, as detailed in [16, 17]. In order to minimize the computational load, we scale down the physical size of the problem (and the grains) by a factor of ~$10^3$. For the simulation, a 1024 nm × 1024 nm × 40 nm geometry was considered and a space discretization size of 4 nm was applied in *x*-and *y*-directions. Note that no periodic boundary condition was applied in order to take into account the realistic effect of demagnetization field in the finite-sized samples. The grains were constructed using a Voronoi Tesselation algorithm. Briefly, a voronoi tessellation algorithm takes randomly chosen finite set of points $p_k$ in a plane and a typical voronoi cell is then constructed around each point (or 'seed') $p_k$ by every point in the plane for which $p_k$ is the nearest seed. Therefore, one can vary the sizes of the grains by simply varying the number of seeds. In our simulation, the low carbon furnace-annealed sample $S_{FA}^{0.05\%}$ is represented as an assortment of 40 nm mean-sized grains and water-quenched sample $S_{WQ}^{0.05\%}$ as 10 nm mean sized grains (Fig. 5(a) and (b) in manuscript respectively). The ratio of these grain sizes are comparable to the experimentally obtained ratio of mean grain-sizes from EBSD.

The exchange energy (*A*) is considered constant within each grain (20 $pJ.m^{-1}$) [18] and allowed to vary from grain to grain randomly within a tolerance of 10% (i.e., 20 $\pm$ 2 $pJ.m^{-1}$). The cubic anisotropy energy is similarly considered constant within a single grain and allowed to vary from grain to grain randomly within a tolerance of 10% (0.10 $\pm$ 0.01 $MJ.m^{-3}$) [19] . For each grain, the cubic anisotropy axis is considered to be randomly oriented, to replicate the randomness in crystallographic distribution of the grains (as seen in Fig. 1(b)-(d)), which in turn alters the effective magnetic anisotropy in the direction of the applied field in each grain. The saturation magnetization ($M_S$) used in the simulation is 400 $kA.m^{-1}$, obtained from experimental data and is assumed homogeneous throughout the geometry.

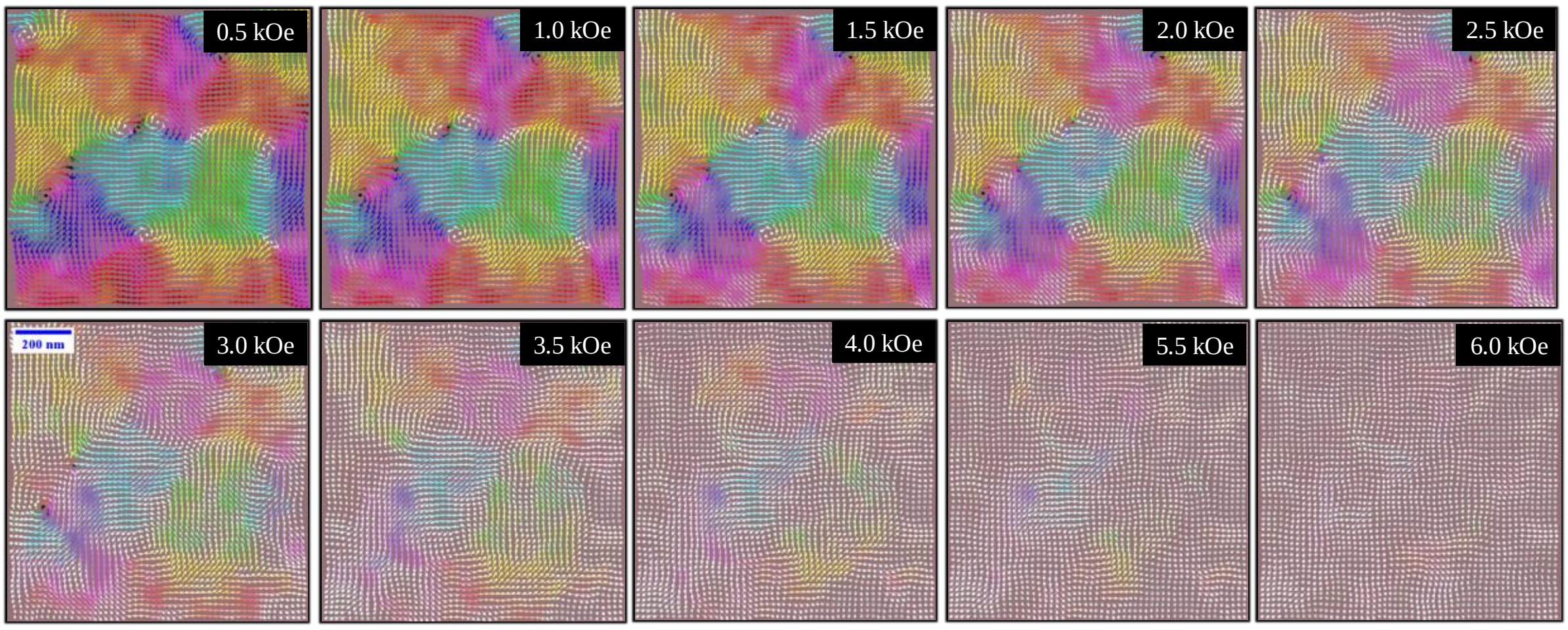

FIG S8. Evolution of the simulated magnetic landscape ($\vec{M}(x, y)$) of the 40 nm grain-sized steel sample (i.e., Fig. 5(a) in main text) as a function of field (0.5-6.0 kOe) applied perpendicular to the plane. See color wheel in Fig. 6 in main text for magnetization directions.

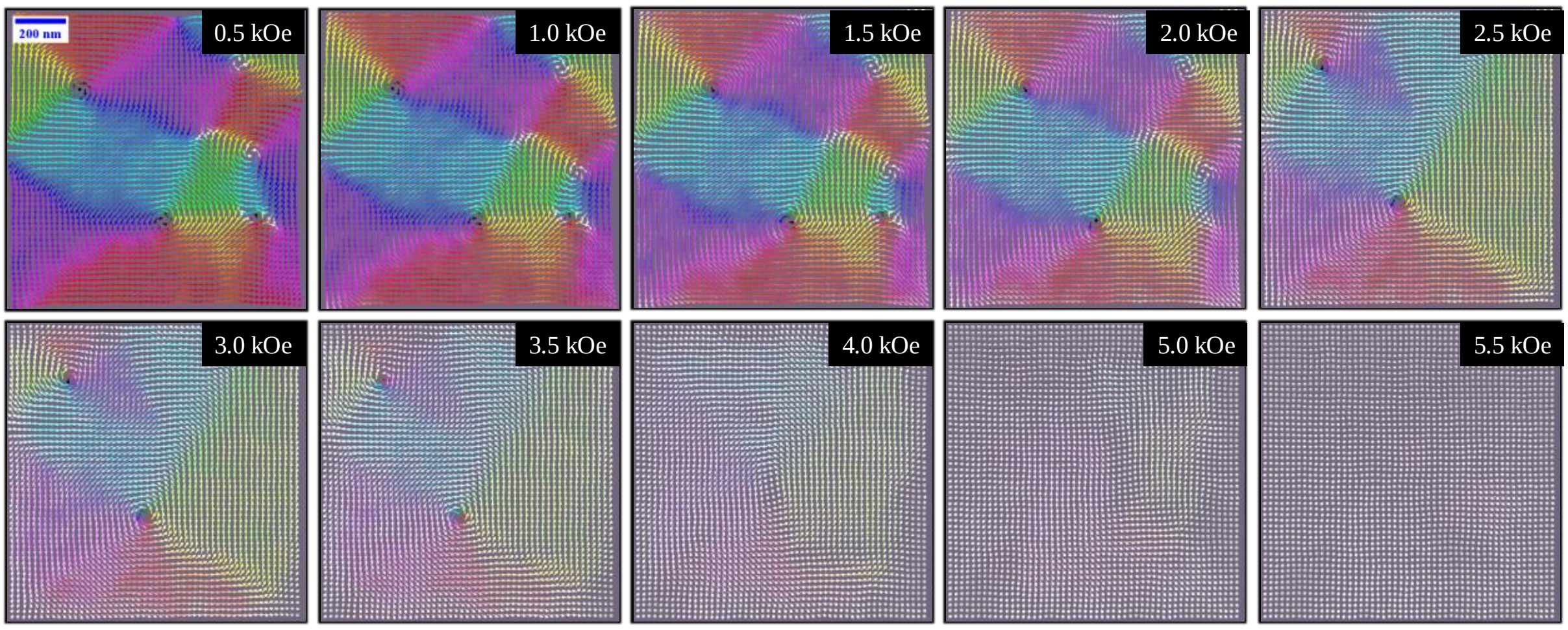

FIG S9. Evolution of the simulated magnetic landscape ($\vec{M}(x, y)$) of the 10 nm grain-sized steel sample (i.e., Fig. 5(b) in main text) as a function of field (0.5-5.5 kOe) applied perpendicular to the plane. See color wheel in Fig. 6 in main text for magnetization directions.

## Supplementary Information – 9

Spatial maps of magnetization on the steel samples were obtained by MFM (Asylum Research, Oxford Instruments). MFM experiments were performed in an ac-tapping mode using a Co/Cr coated magnetic tip with typical resonant frequency ~71 kHz. For a typical MFM experiment, the tip is magnetized in a direction normal to the sample surface and the magnetized tip then raster-scans the sample and the surface-topography is measured. A similar line-scan is done again with a constant lift height of the tip. This lift, typically in tens of nm, results in excluding the short-range atomic forces and takes into account the long-range electromagnetic forces exclusively. Note that the lift-height (i.e. the tip-sample distance) is a crucial parameter in MFM experiments [20]. We acquired MFM images with varying lift heights and optimized the parameter to be ~ 30 nm for our experiments. The magnetic force on the tip, $\vec{F} = \mu_0 \vec{\nabla}(\vec{m}.\vec{H})$ which can be attractive or repulsive (depending on the relative orientation of the dipoles in the tip and the sample, results in a shift in the phase ($\varphi$) and frequency ($f$) of the cantilever oscillation ($\Delta\varphi, \frac{\Delta f}{f} \propto \frac{\partial F_z}{\partial z}$). Thus, the magnetic texture is revealed through changes in these quantities. Figs. 6(d)-(f) in manuscript show the dual-pass phase images of the $S_{AR}^{0.05\%}$, $S_{FA}^{0.05\%}$ and $S_{WQ}^{0.05\%}$ steel samples, respectively in the demagnetized state ($H$ = 0 Oe) for a tip-sample distance of 30 nm (see the color map denoting $\Delta\varphi$ adjacent to each image).

The surface topography of the polished steel samples were imaged by ac-tapping mode Atomic Force Microscopy (AFM) measurement using a Al(100) coated Si probe-tip with typical frequency ~70 kHz and spring constant 2 N/m in an AC240TS-R3 model Asylum Research AFM.